\documentclass{aa}  

\usepackage{graphicx}
\usepackage{txfonts}
\usepackage[colorlinks=true,allcolors=blue]{hyperref}
\usepackage{amsmath}	
\usepackage{amsmath,amssymb}
\usepackage{mathrsfs}
\usepackage[flushleft]{threeparttable}
\usepackage{siunitx}
\usepackage{xcolor}
\usepackage{booktabs, caption}
\usepackage{float}
\usepackage{pdflscape}
\usepackage{ulem}
\usepackage{physics}
\usepackage{natbib}
\bibpunct{(}{)}{;}{a}{}{,} 
\begin{document}

    \title{A Detailed Look at the Most Obscured Galactic Nuclei in the Mid-Infrared}

   \author{F. R. Donnan\inst{1}, D. Rigopoulou\inst{1}, I. Garc{\'i}a-Bernete\inst{1}, M. Pereira-Santaella\inst{2}, A. Alonso-Herrero\inst{3}, P. F. Roche\inst{1},  S. Aalto\inst{4}, A. Hern\'an-Caballero\inst{5}, H. W. W. Spoon\inst{6}}
   \authorrunning{F. R. Donnan et al.}

   \institute{$^{1}$ Department of Physics, University of Oxford, Keble Road, Oxford, OX1 3RH, UK\\
              \email{fergus.donnan@physics.ox.ac.uk}\\
              $^{2}$ Observatorio Astron\'omico Nacional (OAN-IGN)-Observatorio de Madrid, Alfonso XII, 3, 28014, Madrid, Spain\\
              $^{3}$ Centro de Astrobiolog\'{\i}a (CAB), CSIC-INTA, Camino Bajo del Castillo s/n, E-28692 Villanueva de la Ca\~nada, Madrid, Spain\\
              $^{4}$ Department of Space, Earth and Environment, Chalmers University of Technology, Onsala Space Observatory, 439 92 Onsala, Sweden\\
              $^{5}$ Centro de Estudios de F\'isica del Cosmos de Arag\'on (CEFCA), Plaza San Juan, 1, E-44001 Teruel, Spain \\
              $^{6}$ Cornell Center for Astrophysics and Planetary Science, Ithaca, NY 14853, USA}

   \date{Received September 15, 1996; accepted March 16, 1997}

 
  \abstract 
  {Compact Obscured Nuclei (CONs) are an extreme phase of galaxy evolution where rapid supermassive black hole growth and$/$or compact star-forming activity is completely obscured by gas and dust.}
   {We investigate the properties of CONs in the mid-infrared and explore techniques aimed at identifying these objects such as through the equivalent width (EW) ratios of their Polycyclic Aromatic Hydrocarbon (PAH) features.}
   {We model Spitzer spectra by decomposing the continua into nuclear and star-forming components from which we then measure the nuclear optical depth, $\tau_N$, of the $9.8 \mu$m silicate absorption feature. We also use Spitzer spectral maps to investigate how PAH EW ratios vary with aperture size for objects hosting CONs.}
   {We find that the nuclear optical depth, $\tau_N$, strongly correlates  with the HCN-vib emission line in the millimetre for CONs with a Pearson correlation coefficient of 0.91. We find the PAH EW ratios technique to be effective at selecting CONs and robust against highly inclined galaxies where strong dust lanes may mimic a CON like spectrum by producing a high $\tau_N$. Our analysis of the Spitzer spectral maps showed that the efficacy of the PAH EW ratios to isolate CONs is reduced when there is a strong star-forming component from the host galaxy. In addition, we find that 
   the use of the inferred nuclear optical depth is a reliable method to identify CONs in $36^{+8}_{-7}\%$ of ULIRGs and $17^{+3}_{-3}\%$ of LIRGs, consistent with previous work.}
   {We confirm mid-IR spectra to be a powerful diagnostic of CONs where the increased sensitivity of JWST will allow identification of CONs at cosmic noon revealing this extreme but hidden phase of galaxy evolution.}


   \maketitle
%
\section{Introduction}

Recent observational evidence has revealed the presence of highly compact ($< 100$ pc), dusty nuclei, in a subset of luminous ($10^{11} L_{\rm IR} < L_{\odot} 10^{12} L_{\odot}$) and ultraluminous ($L_{\rm IR} > 10^{12} L_{\odot}$) infrared galaxies (LIRGs and ULIRGs respectively) \citep[e.g.][]{Aalto2015, Falstad2021}. These objects represent a crucial phase in galaxy evolution where the nucleus builds mass rapidly through 
star formation and/or super-massive black hole (SMBH) growth \citep[e.g.][]{Garcia-Burillo2012, Gao2021}. As SMBHs are known to co-evolve with their host galaxy \citep[e.g.][]{Kormendy2013}, these nuclei may play a key role in this co-evolution.

Compact obscured nuclei (CONs) exhibit high column densities of gas (N$_{\rm H} > 10^{25} \rm{cm}^{-2}$) and dust which are highly opaque at short wavelengths (UV/Optical). This has prevented the nature of the power source from being uncovered. It is unclear whether the nuclear power source is accretion onto a SMBH and/or a compact nuclear starburst with a top-heavy IMF \citep[e.g.][]{Hickox2018, Tacchella2018, Kocevski2017}.

Rotational-vibrational lines of HCN have been used to identify and probe the radiation field hidden behind the dust \citep[e.g.][]{Aalto2015, Aalto2019, Falstad2021}. These lines are a result of a heat trapping effect \citep[e.g.][]{Kaufman1998, Rolfss2011, Gonzalez-Alfonso2019} where the extreme density of dust/gas traps radiation from a central heating source, increasing its internal temperature and populating the vibrational states of HCN while the external dust remains cool. \citet{Falstad2021} carried out a search for CONs in the local universe ($z<0.08$) using a representative sample of 46 galaxies, finding CONs more frequently in ULIRGs than LIRGs. However this technique relies on a relatively weak emission feature which is extremely challenging to detect at higher redshifts, prompting the need for alternate techniques.

\citet{Garcia-Bernete2022} showed that 
Polycyclic Aromatic Hydrocarbons (PAH) features in the mid-infrared can be used to detect highly obscured nuclei through measuring their equivalent width (EW) ratios. The presence of a highly obscured nucleus will cause deep absorption at 9.8 $\mu$m from silicates which will suppress the nuclear continuum around the 11.3 $\mu$m PAH feature, increasing its EW relative 
to the 6.2 $\mu$m and 12.7 $\mu$m PAH features that 
are less affected by extinction. Therefore highly obscured nuclei can be revealed through lower EW ratios of the 12.7 $\mu$m feature relative to the 11.3 $\mu$m and the 6.2 $\mu$m to the 11.3 $\mu$m PAH \citep[][]{Garcia-Bernete2022}. Pure star-forming galaxies show near constant PAH EW ratios as the intrinsic flux ratio is approximately constant and the PAHs are subject to the same extinction as the continuum \citep[][]{Hernan-Caballero2020}.

Additional techniques of identifying
highly obscured nuclei in the mid-infrared include
the 14 $\mu$m HCN absorption feature \citep[][]{Lahuis2007} or the various crystalline absorption bands \citep[e.g.][]{IDEOS}.
However, the former was difficult to observe with Spitzer as this feature is faint and easily diluted while the latter relies on absorption at 28 $\mu$m and 33 $\mu$m which are outwith the observing range of JWST, limiting future application. 

In this work we investigate the properties of deeply obscured nuclei based on mid-infrared spectroscopy and evaluate  the efficacy of
different methods of identifying such objects. Utilising Spitzer IRS \citep[The Infrared Spectrograph,][]{Houck2004} spectra, we introduce a new spectral decomposition method which splits the continuum into a relatively unobscured star-forming component and a nuclear component  which is subject to higher extinction. We apply the new decomposition method to samples of LIRGs and ULIRGs to determine the properties of their nucleus. We also use spatially resolved spectra to test the how dilution from the host galaxy affects the measured PAH EW ratios.

The paper is structured as follows: in Section \ref{sec:Obs} we describe the samples of the IRS staring mode data and the construction of cubes for the spectral mapping data. In Section \ref{sec:SpectralFitting} we describe our new spectral decomposition method. In Section \ref{sec:Results} we explore the properties derived from the spectral decomposition and analyse the spectral mapping data. Finally in Section \ref{sec:Discussion} we estimate the number of CONs in ULIRGs and LIRGs, discuss the effect of galaxy inclination and prospects for future work using the James Webb Space Telescope.

\section{Observations}
\label{sec:Obs}
In this work we use Spitzer spectroscopy for a number of galaxy samples. The majority of the data presented are low-resolution staring mode spectra. The reduced IRS staring mode spectra were obtained from the Infrared Database of Extragalactic Observables from Spitzer \citep[IDEOS,][]{IDEOS}.

\subsection{Staring Mode Data}

For the present study we used two main samples of 
LIRGs and ULIRGs.
For a representative sample of  ULIRGs we use the HERschel Ultra Luminous Infrared Galaxy Survey sample \citep[HERUS, ][]{Farrah2013} which consists of the 42 local ULIRGs that were observed with Spitzer. For a complete sample of LIRGs we use the Great Observatories All-sky LIRG Survey \citep[GOALS,][]{Armus2009} sample which consists of 179 LIRGs and 22 ULIRGs. As not all of these have been observed with Spitzer, this brings the sample down to 143 LIRGs and 15 ULIRGs.

We also used a reference sample of purely star-forming galaxies from \citet{Hernan-Caballero2020} where full details of the sample selection can be found. 
In addition we used the CONquest sample which consists of 44 declination and distance limited objects selected from the IRAS revised bright galaxy sample \citep[][]{Sanders2003}. Full details can be found in \citet{Falstad2021}. Of these 44 galaxies, 29 have Spitzer spectra with the majority of the objects without spectra being sub-LIRGs. 

\subsection{Spectral Mapping Data}
In addition to the staring mode spectra we also employed a sub-sample of LIRGs from the GOALS sample that had Spitzer IRS spectral mapping observations with the SL1 and SL2 modules, providing spectra between $\sim 5 - 14$ $\mu$m \citep[][]{Alonso-Herrero2009, Pereira-Santaella2010, Alonso-Herrero2012}. We chose targets without staring mode spectra with the exception of NGC 6926, which was selected as a CON candidate by \citet{Garcia-Bernete2022} to provide an additional candidate to investigate. We also excluded a few objects (IRAS 08339+6517, NGC 6052, NGC 6240, NGC 6621) that had poor signal to noise ratio or issues with the data reduction. The resulting sample presented in this work is shown in Table \ref{tab:Sources} with some basic properties.

The Basic Calibrated Data (BCD) were obtained from the Spitzer Heritage Archive \footnote{\url{https://sha.ipac.caltech.edu/applications/Spitzer/SHA/}} and subsequently combined to construct spectral cubes using CUBISM \citep{Smith2007b}. This process combines the different slit exposures, subtracts the sky background and rejects bad pixels yielding a spectral cube for each module and an associated cube containing the flux errors.

To accurately extract spectra from the two SL modules, we determine the centre location of the object of interest and spatially align the two modules. We use the DAOStarFinder algorithm \citep{Stetson1987} from the \textsc{photutils} python package, to obtain positions of sources in the image. We apply this technique to integrated intensity (moment 0) maps of the 5.4 $\mu$m continuum (5.47 - 5.44 $\mu$m), to identify the position of the nucleus. In order to ensure the cubes are spatially aligned we generate integrated intensity maps of the overlapping channels (7.53 - 7.6 $\mu$m), and find the position where the flux peaks. The spatial offset between the two modules was generally found to be small but non-negligible in some cases (up to $\sim 1"$). This offset correction was then applied when extracting the spectra from each cube.

\begin{figure*}
\hspace*{1.5cm}                                                           
        \includegraphics[width=15cm]{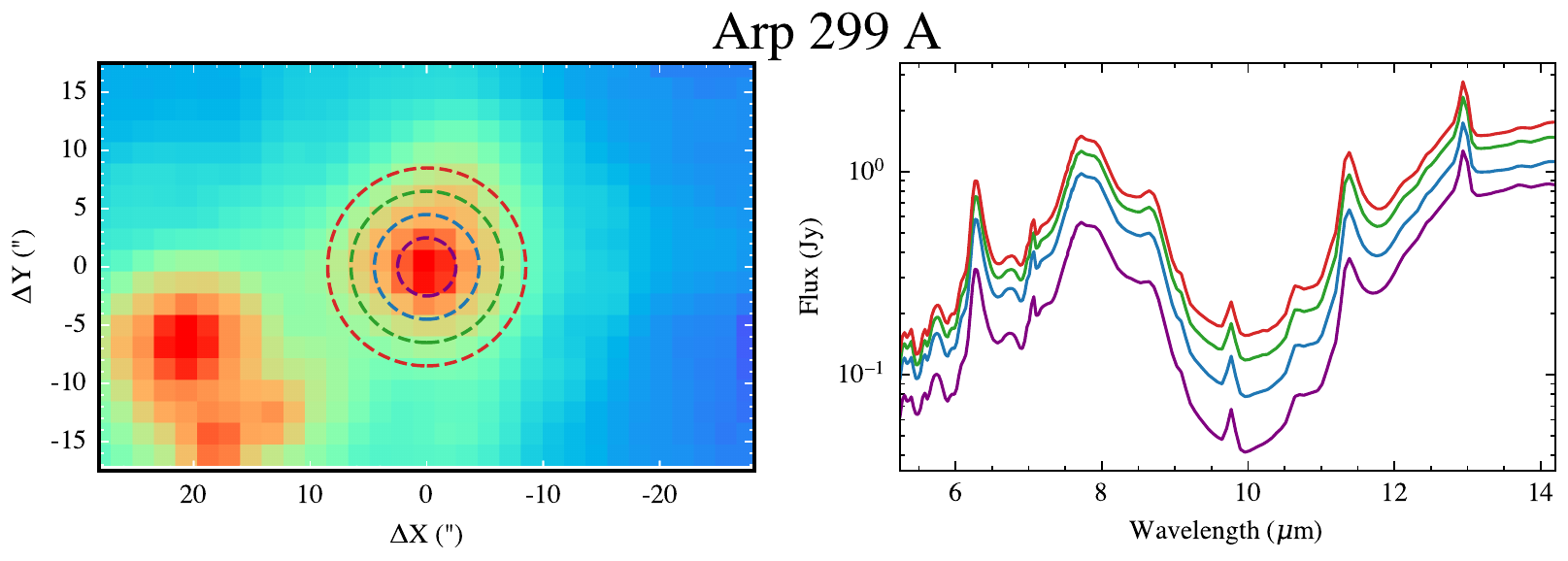}
    \caption{Example of the spectral extraction, in this case for Arp 299 A, at different spatial scales. {\it Left}: The integrated intensity map of the SL1 cube (7.46-14.29 $\mu$m) overlayed with circular apertures of radii 2.5, 4.5, 6.5, 8.5'' shown as the dashed lines. {\it Right}: Extracted spectra from the apertures in the left panel with the corresponding colour. The innermost spectrum (purple) has had an aperture correction applied. } 
    \label{fig:SpecExtract} 

\end{figure*}

As the PSF for these observations is large (FWHM $\sim 3.8"$ at 14 $\mu$m), we extracted spectra in circular apertures each of which provide different dilutions of the galaxy disk with respect to the nuclear region. Specifically these were circular apertures of radius 2.5, 4.5, 6.5, 8.5 arcsec which are shown in Fig. \ref{fig:SpecExtract} for Arp 299 A.

The smallest aperture was corrected for slit losses using a standard star to achieve a point source extraction for the nuclear region. The correction factor was obtained by calculating the flux ratio between an aperture containing the total flux of the star (7.5") and the nuclear aperture (2.5"). A smooth function of the correction factor as a function of wavelength was obtained by fitting a 4\textsuperscript{th} order polynomial to the measured flux ratios. This factor was subsequently applied to the 2.5" spectral extractions.

The majority of spectra analysed in this work consists only of the SL1 \& SL2 modules. This is because the longer wavelength LL1 \& LL2 modules probe larger spatial scales and so galaxies containing an AGN or CON will have different relative contributions of the host at long wavelengths compared to short wavelengths if the full spectral range was used. Therefore, in order to avoid making physical assumptions when applying a correction factor we simply restrict our analysis to the SL modules.

\begin{table*}
\centering
  \caption{Spectral Mapping Sources}
  \label{tab:Sources}
    \def\arraystretch{1.2}
    \setlength{\tabcolsep}{4pt}
    \begin{threeparttable}
  \begin{tabular}{cccccccc}
  
    \hline

    Name & RA  & Dec & $z$ & $\log \left(L_{\rm IR}/L_{\odot}\right)$  & Merger Class \\
    (1) & (2) & (3) & (4) & (5) & (6) \\
    \hline
    Mrk 938 & 00h11m06.56s & -12d06m28.2s & $0.01962$ & 11.49 & d\\
    MCG-02-01-051A  & 00h18m49.85s & -10d21m34.0s & $0.02710$ & 11.48 & b\\
    Arp 236 B (VV 114 W)&  01h07m46.72s & -17d30m27.9s & $0.02007$ & & \\
    Arp 236 A (VV 114 E) &  01h07m47.54s & -17d30m25.6s & $0.02007$ & 11.71 & c &  \\
    UCG 03410 (UGC03405) & 06h13m57.90s & +80d28m34.7s & $0.01247$ & 11.10 & N\\
    ESO557-G002 & 06h31m45.71s & -17d38m44.9s & $0.02099$ & 11.25 & - \\
    NGC 2342 A (NGC2341) & 07h09m12.01s & +20d36m11.2s & $0.01722$ & 11.31 & a\\
    NGC 2369 & 07h16m37.7s & -62d20m37s & $0.01081$ & 11.16 & a &  \\
    NGC 2388 A (NGC2389) & 07h29m04.59s & +33d51m38.0s & $0.01316$ & 11.28 & a\\

    Arp 299 B (NGC 3690) &  11h28m30.987s & 58d33m40.80s & $0.01041$ &  \\
    Arp 299 A (IC 694) &  11h28m33.626s & 58d33m46.65s  & $0.01041$ & 11.93 & c &  \\
    ESO320-G030 & 11h53m11.7s & -39d07m49s & $0.01078$ & 11.17 & N & \\
    NGC 4922 & 13h01m25.27s & +29d18m49.5s & $0.02359$ & 11.38	 & c\\
    MCG-03-34-063 & 13h22m19.02s & -16d42m30.0s & $0.02133$ & 11.28	 & d\\
    NGC 5257 A (NGC5258) & 13h39m57.72s & +00d49m53.0s & $0.02254$ & 11.62 & b \\
    NGC 5395 & 13h58m37.96s & +37d25m28.1s & $0.01158$ & 11.08 & N\\
    IC 4518 E & 14h57m45.33s & -43d07m57.0s & $0.01573$ & 11.23 & a\\
    ZW 049.057 & 15h13m13.1s	& +07d13m32s & $0.01306$ & 11.35 & N & \\

    IC 4687 & 18h13m39.80s & -57d43m30.7s & $0.01734$ & 11.62 & b\\
    NGC 6786 A & 19h11m04.37s & +73d25m32.5s & $0.02502$ & 11.49 & c \\
    NGC 6926 & 20h33m06.1s & -02d01m39s & $0.02001$ & 11.32 & d	\\

    IC 5179 & 22h16m09.1s & -36d50m37s & $0.01141 $ & 11.24	& N & \\
    NGC 7552 & 23h16m10.7s & -42d35m05s & $0.00536$ &11.11 & N\\
    NGC 7592 B (W)& 23h18m21.78s & -04d24m57.0s & $0.02444$ & & \\
    NGC 7592 A (E) & 23h18m22.60s & -04d24m58.0s & $0.02444$ & 11.40	 & b \\
    NGC 7674 & 23h27m56.71s & +08d46m44.3s & $0.02903$ & 11.56 & a\\
    NGC 7752 A (NGC 7753) & 23h47m04.84s & +29d29m00.5s & $0.01720$ & 11.07 & c \\

    \hline
  
  \end{tabular}
\begin{tablenotes}
    \item[] Column (1): Source name. Column(2): Source right ascension from GOALS targets \citep[][]{Armus2009}. Column (3): Source declination from GOALS targets. Column (4): Redshift obtained from NED. Column (5): Total IR luminosity. Many of these contain multiple nuclei - see \citet{Armus2009} for more details. Column (6): Merger Class from \citet{Stierwalt2013} where N = nonmerger, a = pre-merger, b = early stage merger, c = mid-stage merger, and d = late stage merger. 
  \end{tablenotes}
  \end{threeparttable}
 \end{table*}

\section{Spectral Fitting}
\label{sec:SpectralFitting}
There are a number of approaches to modelling the mid-infrared spectrum of galaxies. A popular tool from the Spitzer era is PAHFIT \citep[][]{Smith2007}, which models the continuum with a series of blackbodies subject to extinction and PAH emission features with a series of Drude profiles. The tool is well suited to model pure star-forming galaxies, however it is less well-suited to objects containing highly obscured nuclei and/or AGN. 
Alternative techniques have been used to address this problem such as QUESTFIT \citep[][]{Veilleux2009} and DeblendIRS \citep[][]{Hernan-Caballero2015}, however, both of these assume 
a set of fixed templates for the PAH emission which prevents us from measuring the properties of the PAHs independently for each object.

The presence of a distinctly different nuclear continuum in heavily obscured objects makes it difficult to model these sources with a single extinction component. We therefore create a new model that retains the PAH emission and spectral lines of the PAHFIT tool but models the continuum with two components, a nuclear and a star-forming one, each subjected to different levels of extinction. In what follows we describe the parameterisation of each component in detail.

\subsection{Star-forming Component}
\begin{figure*}
\hspace*{1.7cm}                                                           
	\includegraphics[width=14cm]{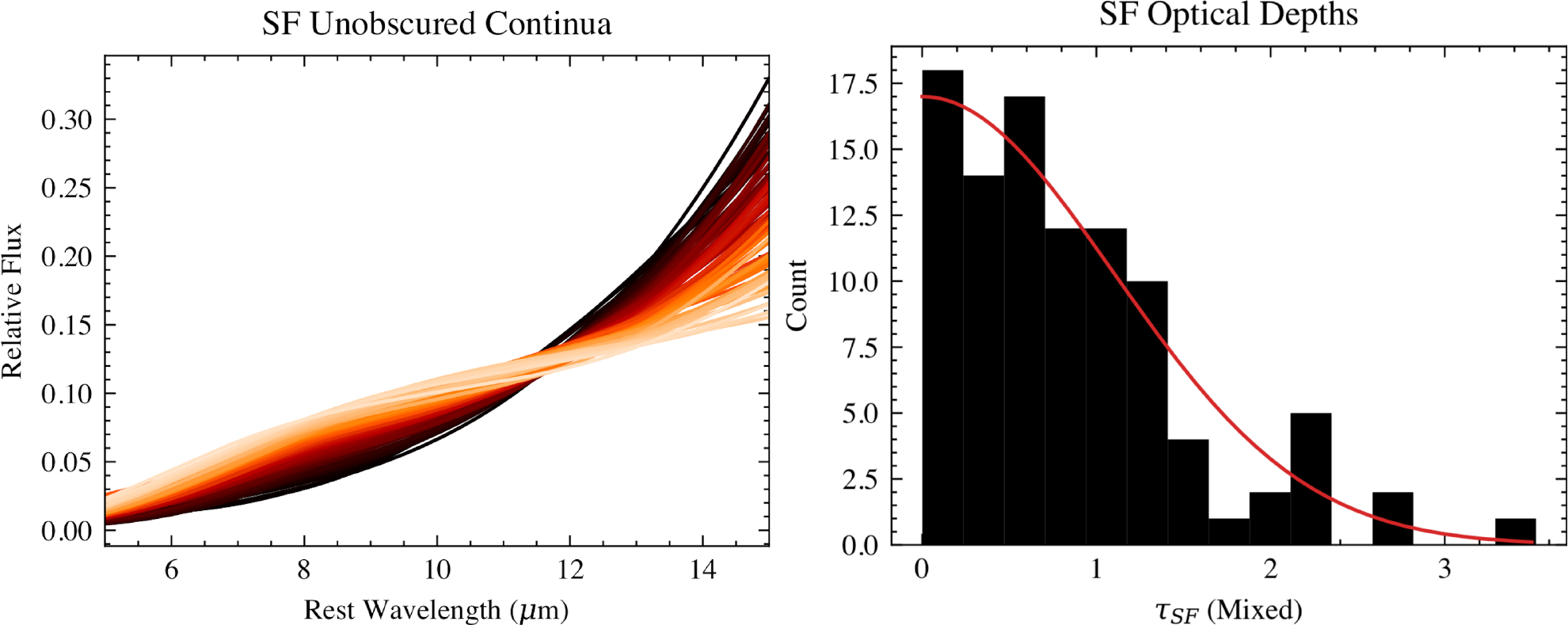}
    \caption{\textit{Left:} Unobscured continua, $C_{\rm SF}(\lambda)$, obtained from the fits to the star-forming sample, used as templates for the spectral decomposition. The colour indicates the steepness (8.0 $\mu$m to 13 $\mu$m) where darker the colour the steeper the continuum. \textit{Right:} Histogram of the optical depths, assuming mixed geometry, for the star-forming sample. The red line shows the half-normal prior used for the spectral decomposition, based on these results.}
    \label{fig:SFTemp} 
\end{figure*}

To model the star-forming component of the continuum we generate templates by fitting the star-forming 
sample of \citep[][]{Hernan-Caballero2020} with PAHFIT. The left panel of Fig. \ref{fig:SFTemp} shows the resulting continuum templates ordered by their flux ratio between 8 and 13 $\mu$m. Rather than selecting an individual template, we linearly interpolate between the templates ordered in this way with a single parameter between 0 and 1 where 0 corresponds to the steepest template. The resulting template, $C_{\rm SF}(\lambda)$, is then subject to extinction (assuming mixed geometry) 
of the form $\frac{1 - e^{-\tau_{\lambda}}}{\tau_{\lambda}}$, where we restrict the optical depth, $\tau_{\lambda}$, using a half-normal prior based on the resulting fits from the star-forming sample. This is shown in the right panel of Fig. \ref{fig:SFTemp}, where the histogram shows the optical depths at 9.8 $\mu$m inferred from the star-forming sample and the red line shows the half-normal prior used for the extinction of the star-forming component for the spectral decomposition.

\subsection{Nuclear Component}
\label{sec:NucComp}
The nuclear component in our model consists of a continuum with some extinction applied. The continuum, $C_{\rm Nuc}(\lambda)$, is modelled as a quadratic spline with three evenly spaced anchor points at 5.5 $\mu$m, 9.85 $\mu$m and 14.2 $\mu$m. The y-value of these knots is allowed to vary as a free parameter. As the continuum is normalised by the total flux and scaled by the nuclear fraction, $\beta$, the spline has effectively only two free parameters. 
Extinction is then applied to the continuum as a screen using the ice template and silicate template derived from NGC 4418 and has the form $e^{-\left(\tau_{\rm{Ice}}(\lambda)  + \tau_{N}(\lambda) \right)}$. The silicate template is normalised to 1 at 9.8 $\mu$m, therefore the optical depth, $\tau_N$, is that measured at 9.8 $\mu$m, although it is worth noting this is an underestimation as this assumes $\tau=0$ at the anchor points when in reality a power law component to the extinction curve causes $\tau>0$ at the anchor points. 

The ice template (including the CH absorption) is allowed to vary separately to the silicate template, resulting in a total of 4 free parameters (2 for the shape of the spline, $C_{\rm Nuc}(\lambda)$, and 2 for the extinction, $\tau_{\rm{Ice}}$ and $\tau_{N}$) determining the shape of the nuclear component. The creation of these templates is discussed in Appendix \ref{sec:AppA}. For the silicate profile we use a template derived from a highly obscured source, namely NGC 4418, because the composition of silicates is likely different (resulting in a different profile) compared to extinction laws derived from the MW \citep[e.g.][]{Tsuchikawa2021} or the line of sight is more complex than towards the galactic centre.

It is also work noting that the optical depth peak of the silicate profile of NGC 4418 is $\sim 4.4$ from the Spitzer spectroscopy whereas ground based observations with a smaller beam find a higher peak $\sim 7$ \citep[][]{Roche2015}. This means that the silicate profile used is still contaminated by some amount of star-formation and so the nuclear optical depth values inferred should be taken as lower limits for the true optical depth. In Appendix \ref{sec:AppD} we compare this silicate template to another one - that of IRAS 08572+3915. From this testing we find the results from this work to be unchanged.

For objects where the nuclear optical depth is minimal, the continuum decomposition is unreliable as there is nothing to differentiate the star-forming component from the nuclear one, however the total continuum should be reliable. Therefore the decomposition is 
particularly sensitive to cases 
where the optical depth is high (as in the case of a CON) and the contribution of the nuclear component is above a certain threshold. We investigate this further in Section \ref{sec:Testing}.

\subsection{Full Model}
The resulting full model is given by 
\begin{multline}
\label{eqn:Model2}
    f_{\nu}(\lambda) = \left[ \sum_{i=1}^{N_{\rm Lines}} I_{\nu, \rm Line}^{(i)} (\lambda) + \sum_{i=1}^{N_{\rm PAH}} I_{\nu, \rm PAH}^{(i)} (\lambda) + (1 - \beta)C_{\rm SF}(\lambda)  \right] \frac{1 - e^{-\tau_{\lambda}}}{\tau_{\lambda}} \\ + \beta C_{\rm Nuc}(\lambda)   e^{-\left(\tau_{\rm{Ice}}(\lambda)  + \tau_{N}(\lambda) \right)},
\end{multline}
where $C_{\rm SF}(\lambda)$ is the star-forming unobscured continuum, $\beta$ is the nuclear fraction, and $C_{\rm Nuc}(\lambda)$ is the nuclear unobscured continuum. These unobscured continua multiplied by their respective extinction factors are normalised by their area to ensure that the scale factor, $\beta$, gives the fractional contribution of the nucleus to the total continuum flux (between 5 - 14 $\mu$m). The area is calculated numerically during the fitting process.

To aid in the decomposition, we enforce a prior on the ratio of the total PAH flux, $f_{\rm PAH}$, to the integrated continuum flux of the star-forming continuum, $f_{\rm SF}$, between 5 - 14 $\mu$m. From the PAHFIT results for the star-forming calibration sample we found this ratio to be $f_{\rm PAH}/f_{\rm SF} \approx 1.92$. We use a very wide normal prior with a standard deviation of 10.0. This wide prior is designed to discourage solutions with zero contribution from the star-forming component as the large apertures of Spitzer will contain at least some extended contribution. The large width of this prior was chosen arbitrarily such that the prior discourages zero star-forming component where $f_{\rm PAH}/f_{\rm SF} \xrightarrow{} \infty$ but does not heavily bias the results.

\begin{figure*}
\hspace*{0.5cm}                                                           
	\includegraphics[width=17cm]{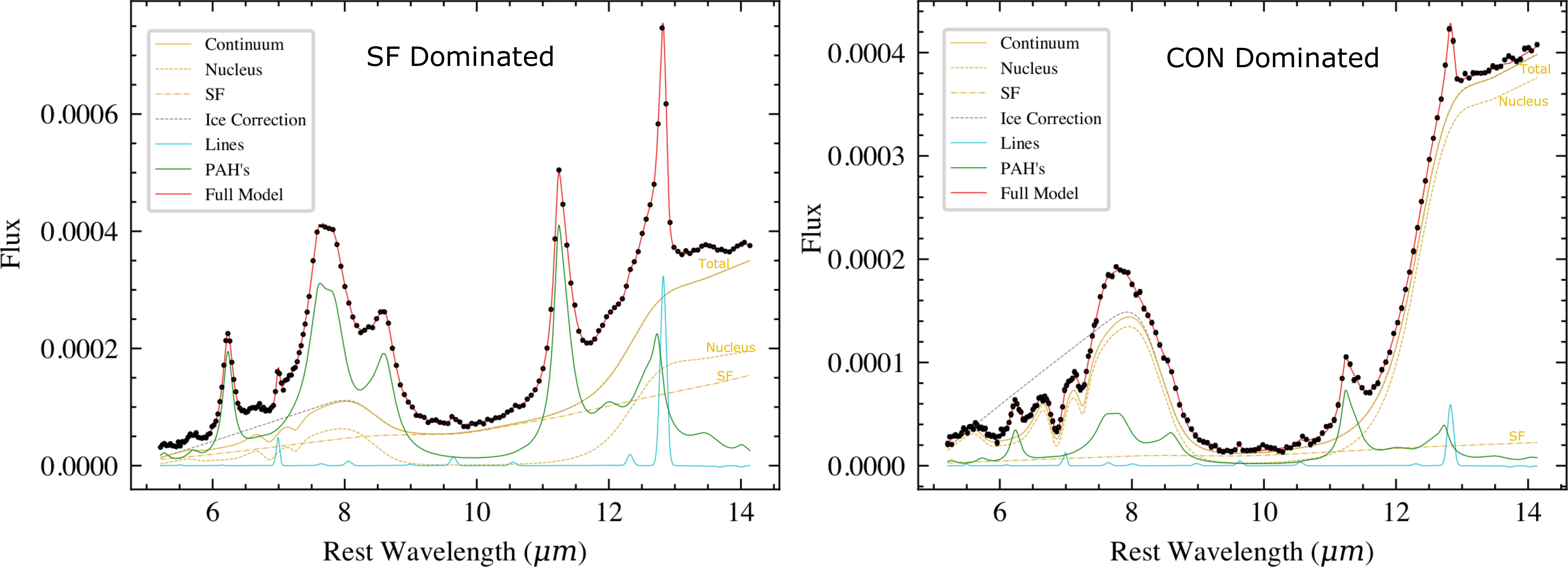}
    \caption{Example of our spectral decomposition fit (described in Section \ref{sec:SpectralFitting}) to simulated data of a galaxies hosting a CON. The left panel shows a highly diluted nucleus ($\beta = 0.42$) with a spectrum dominated by star formation and the right panel shows a CON dominated source ($\beta = 0.89$). The full model is shown in red composed of the various components shown in the legend. The continuum is shown as the solid gold line which is a sum of the nuclear and star-forming (SF) components shown as the two dashed lines. The flux is in arbitrary units. The equation of the full model is given in Equation (\ref{eqn:Model2}).}
    \label{fig:DecompModel} 
\end{figure*}

\subsection{Bayesian Inference}
\label{sec:ModelFitting}

To fit these models to the spectra, we use a Bayesian approach to provide accurate posterior probabilities for the galaxy properties of interest. We use MCMC sampling from \textsc{numpyro} \citep{Phan2019} to sample the relative posterior probability of the model given the data, $\textrm{Pr}(M|D)$, which is given by 
\begin{equation}
\label{eqn:prob}
    \ln{\textrm{Pr}(M|D)}= \ln{\textrm{Pr}(M)} + \ln{\textrm{Pr}(D|M)} + \rm{const}
\end{equation}
where the prior, $\textrm{Pr}(M)$, is uniform between sensible limits as discussed in the previous sections and the log-likelihood, $\ln{\textrm{Pr}(D|M)}$, is effectively just the chi-squared as the error bars are fixed by the data:
\begin{equation}
\label{eqn:chi2}
    \ln{\textrm{Pr}(D|M)} = \sum_{i=1}^{N}\left( \frac{\left(f_i - f_{\nu}(\lambda) \right)^2}{\sigma_i^2} \right) +\rm{const}
\end{equation}
where there are $N$ flux data $f_i$, with error $\sigma_i$. From \textsc{numpyro}, we use the No-U-Turn Sampling (NUTS), a Hamiltonian Monte Carlo method which allows for faster sampling through parallelisation compared to traditional MCMC methods. After a burn-in of 2000 samples we take another 2000 samples to obtain marginalised posteriors for each of the parameters. 

From the MCMC samples, posterior probability distributions for the properties derived from the spectra can be constructed. We are primarily interested in the EW of the PAH features, which are calculated numerically for each sample of the parameters using
\begin{equation}
\label{eqn:eqw}
    \rm{EW} = \int \frac{f^{\rm PAH}_{\nu}}{f^{\rm cont}_{\nu}}  d\lambda , 
\end{equation}
where $f_{\nu}^{\rm PAH}$ is the PAH profile and $f_{\nu}^{\rm cont}$ is the continuum. The equivalent width ratios of the various PAH features can then be calculated with associated error bars. The continuum used is the ice-corrected continuum to allow comparisons with the IDEOS sample \citep{IDEOS}, and to better determine how the silicates may affect the PAH features. The integrated flux of the PAH features are also calculated using 
\begin{equation}
\label{eqn:flux}
    f^{\rm PAH}  = \int f_{\nu}^{\rm PAH} d\nu,
\end{equation}
where posteriors for the PAH flux ratios can be calculated. We do this with and without extinction applied to obtain estimates for the intrinsic ratios as well.

We also estimate the strength of the 9.8 $\mu$m silicate feature defined as \citep{Spoon2007}
\begin{equation}
\label{eqn:SilStrength}
    S_{\rm{sil}} = \ln\left(\frac{f_{9.8, \rm{obs}}^{\rm cont}}{f_{9.8, \rm{int}}^{\rm cont}}\right),
\end{equation}
which is the log of the ratio of the observed to the intrinsic continuum at 9.8 $\mu$m. Note that this is different to the nuclear optical depth, $\tau_{N}$, in the spectral decomposition method, which allows the nuclear silicate feature to be ``filled-up'' by the star-forming dust continuum. Therefore, objects with a buried nucleus but with a strong extended star-forming component may have a low observed silicate depth as defined by Equation \ref{eqn:SilStrength} but a deep nuclear optical depth, $\tau_{N}$.

\subsection{Testing the Method}
\label{sec:Testing}
The primary motivation for modelling the spectra using this new decomposition tool is to provide a realistic continuum to enable us to 
accurately infer the properties of the emission features. However it also allows us to 
better constrain the physical properties of the two components (nuclear and star-forming), 
particularly the optical depth of the nuclear region, the fraction of the nuclear to the total continuum and, consequently, the shape of the nuclear continuum. It is therefore instructive to test the effectiveness/robustness of this decomposition method before inferring the properties of the deeply obscured galaxies studied here.

\subsubsection{Accuracy of Inferred Properties}
We first generate simulated data of a typical star-forming galaxy hosting a CON (with a given $\tau_N$) with varying degrees of dilution from the host galaxy (different values of $\beta$). This is done by constructing a spectrum using a star-forming galaxy and a CON template,
where different fractional contributions, $\beta$, of the CON are used. We then 
run the decomposition tool on the spectra
and test its ability to recover the true nuclear fraction and nuclear optical depth. 

We use NGC 1797 for the star-forming component as this shows a typical spectrum of a star-forming galaxy with some silicate absorption, and the nucleus of NGC 4418 as the CON template. The star-forming continuum is normalised to the integrated continuum flux between 5 and 14 $\mu$m. The PAH and emission lines from NGC 1797 are also included as part of the star-forming component and are normalised by the same factor as the continuum. The CON continuum from NGC 4418 is normalised by its integrated flux over the same wavelength range. The star-forming component is then scaled by $(1-\beta)$ and the CON component by $\beta$ for 20 values of $\beta$ between 0 and 1. We generate two sets of simulated data, the first with a CON component with $\tau_N = 4.5$ and the second with a lower nuclear optical depth of $\tau_N = 3.0$ to test the reliability of the method for different levels of nuclear obscuration.

\begin{figure}
	\includegraphics[width=\columnwidth]{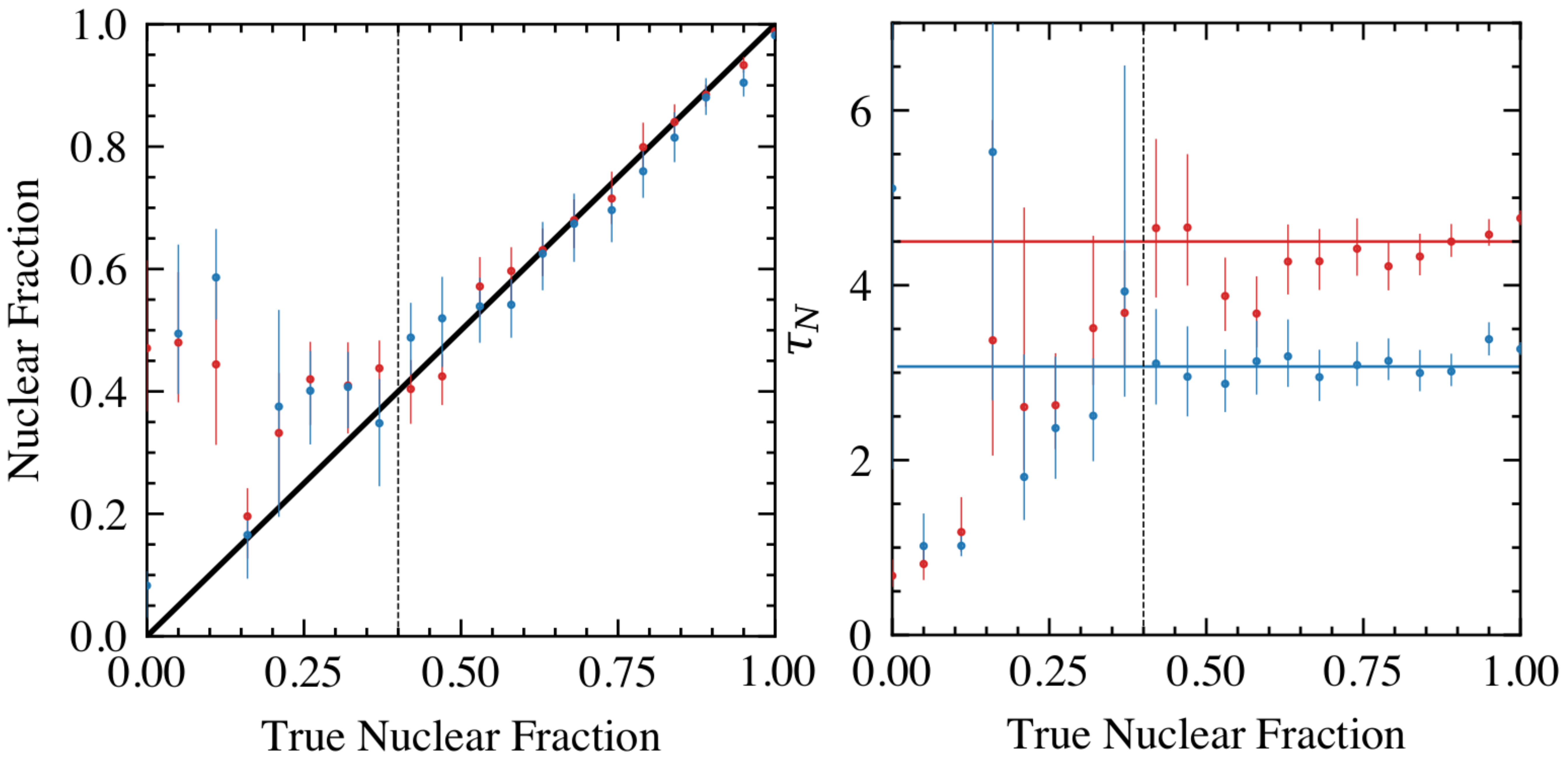}
    \caption{Testing the recovered properties of simulated data using the spectral decomposition fitting method. {\textit Left:} Measured nuclear fraction, $\beta$, against the true value. The solid black line shows where the measured equals the true value. {\textit Right:} Measured nuclear optical depth, $\tau_N$, against the true nuclear fraction. The solid horizontal lines show the true optical depth of $\tau_N = 4.5$ in red and $\tau_N = 3.0$ in blue. The vertical dashed black lines show a cutoff of $\beta = 0.4$ below which the measured values become unreliable. }
    \label{fig:TestFig} 
\end{figure}

Fig. \ref{fig:DecompModel} shows two example fits to the simulated data with the left panel showing a CON highly diluted by the host galaxy ($\beta = 0.42$) and the right shows a CON dominated source ($\beta = 0.89$). In both cases the nuclear optical depth, $\tau_N$, and nuclear fraction, $\beta$, are successfully recovered by the model. At values of $\beta \lesssim 0.4$, we find that the method is unable to recover the nuclear fraction and$/$or the optical depth as reliably, although in the case of the optical depth, the errors do a reasonably good job of accounting for this. This is demonstrated in Fig. \ref{fig:TestFig}, where the measured nuclear fraction is plotted against the true value of the nuclear fraction in the left panel. The red points shown the data set with a $\tau_N = 4.4$ whereas the blue show a lower $\tau_N = 3.0$. This plot demonstrates that even when the nuclear component has a lower optical depth, the nuclear component is still recovered by the model. The right panel shows the measured optical depth for each simulated spectrum with the true value of $\tau_N = 4.5$ displayed for the red points and $\tau_N = 3.0$ for the blue points. 

In the cases of maximum dilution $\beta \lesssim 0.15$, the optical depth recovered is significantly below the true value but with a high measured value of $\beta$. This is likely due to the nuclear component being used to fit the star-forming component as the two components become degenerate when $\beta$ is low. We therefore conclude that the optical depth is accurately recovered for values of $\tau_{N} > 2.5$ ($> 5\sigma$ of the silicate depths of the star-forming calibration sample) and nuclear fractions of $\beta > 0.4$.

\subsubsection{Multiple Apertures Fitting}
From the spectral mapping data, each aperture yields a spectrum (see Fig. \ref{fig:SpecExtract}) that we would like to fit. As each of these apertures has the same centre but different radii they contain the same nucleus but different star-forming fractions.
It is therefore important to fit all these spectra simultaneously assuming the same nuclear component for all the spectra but different star-forming components for each.

To test this idea, we use the spectral map of a known CON to measure the nuclear properties from the spectra of each aperture individually and then compare the results to those from fitting all the apertures simultaneously. We use ESO 320-G030 which is known to be a CON but has a strong star-forming component \citep[e.g.][]{Alonso-Herrero2006, Gonzalez-Alfonso2021}.
\begin{figure}
	\includegraphics[width=\columnwidth]{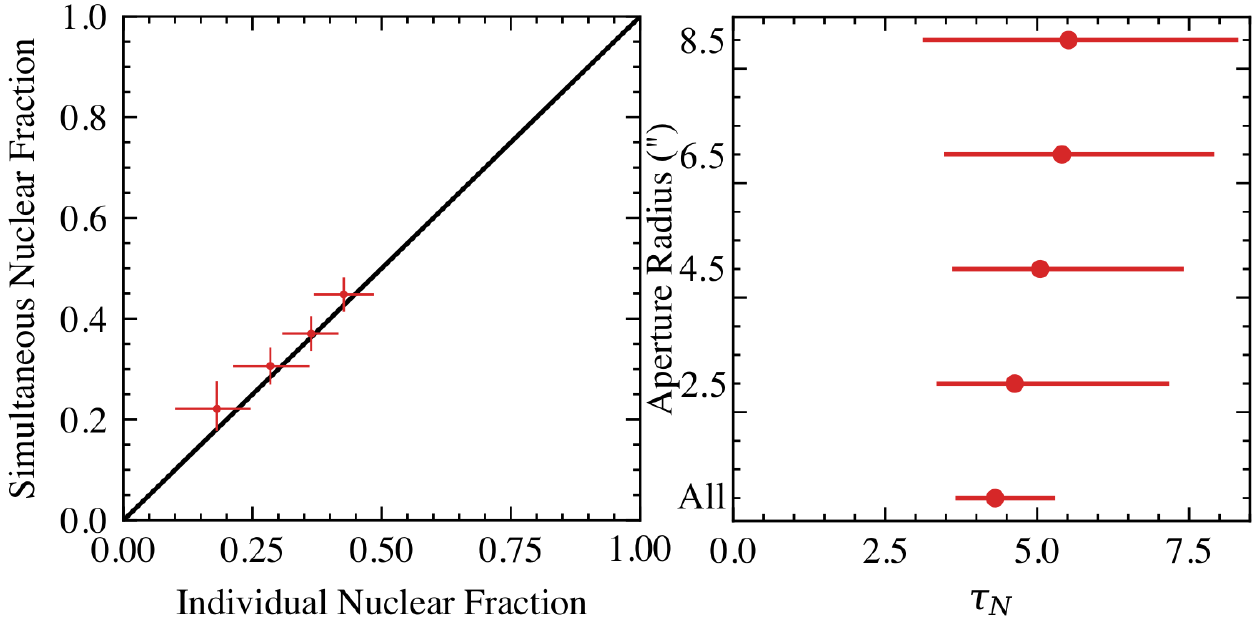}
    \caption{Comparison between fitting spectra from individual apertures vs fitting all apertures simultaneously with shared nuclear parameters for ESO320-G030. {\textit Left:} Nuclear fraction, $\beta$, measured from the simultaneous fitting on the y-axis vs the individual spectra on the x-axis. The solid black line shows where these are equal. {\textit Right:} Measured optical depth of the nucleus for each aperture individually and for all apertures simultaneously. Note the much smaller error when using all the spectra simultaneously. }
    \label{fig:ESO320Test} 
\end{figure}
The left panel of Fig. \ref{fig:ESO320Test} compares the measured nuclear fraction for each aperture when fitting all the spectra simultaneously sharing the same nucleus against fitting each of the spectra independently. The measured values are consistent within 1$\sigma$ between the two methods. The right panel of Fig. \ref{fig:ESO320Test} shows the measured optical depth of the nucleus measured from each aperture individually compared to those measured from all apertures when fit simultaneously. 
Again, we find that the measurements are consistent between the two methods and crucially, even in the presence of significant contribution from the host galaxy.
In this case we find that the value of $\tau_N$ does not change, instead the errorbars expand accordingly. This test also demonstrates the added value in fitting all the apertures simultaneously, with tighter constraints on the nuclear optical depth. Having proved the robustness of our method, we proceed and deploy the tool to fit the spectral mapping data.

\section{Results}
\label{sec:Results}
Based on the spectral fits using the new decomposition tool, we first refine the CON selection criteria (Section \ref{sec:SelectionCrit}) before investigating the physical properties of the nucleus in Section \ref{sec:CONDef} and Section \ref{sec:SpecShape}. We then present results using the spectral mapping data in Section \ref{sec:SpecMappingResults}.

\subsection{Refining the CON selection criteria}
\label{sec:SelectionCrit}
To select CONs using PAH EW ratios, \citet{Garcia-Bernete2022} defined two ratios, the EW(12.7)/EW(11.3) and EW(6.2)/EW(11.3). Both of these are the ratio of the EW of a PAH feature outwith the 9.8 $\mu$m silicate absorption (12.7 $\mu$m and 6.2 $\mu$m) relative to the 11.3 $\mu$m PAH which is located  within this absorption band. Using the star-forming sample of \citet{Hernan-Caballero2020}, \citet{Garcia-Bernete2022} showed that both of these ratios were constant for the sample although the 6.2/11.3 ratio shows a larger scatter. 
As the presence of a heavily obscured nucleus results in a lower continuum around the 11.3 $\mu$m PAH feature compared to the 6.2 $\mu$m or 12.7 $\mu$m PAH, the CON selection region was defined by values of the EW(12.7)/EW(11.3) and EW(6.2)/EW(11.3) ratios below those found for the star-forming sample.

As the previous PAH EW thresholds for selecting highly obscured nuclei from \citet{Garcia-Bernete2022} were based on IDEOS fits using Pearson-IV profiles for the PAH features, we need to 
refine those criteria in line with our new fitting method. 
For this purpose we therefore re-fit the \citet{Hernan-Caballero2020} star-forming sample using our new spectral decomposition method.

\begin{figure*}
\hspace*{1.cm}                                                           
	\includegraphics[width=16cm]{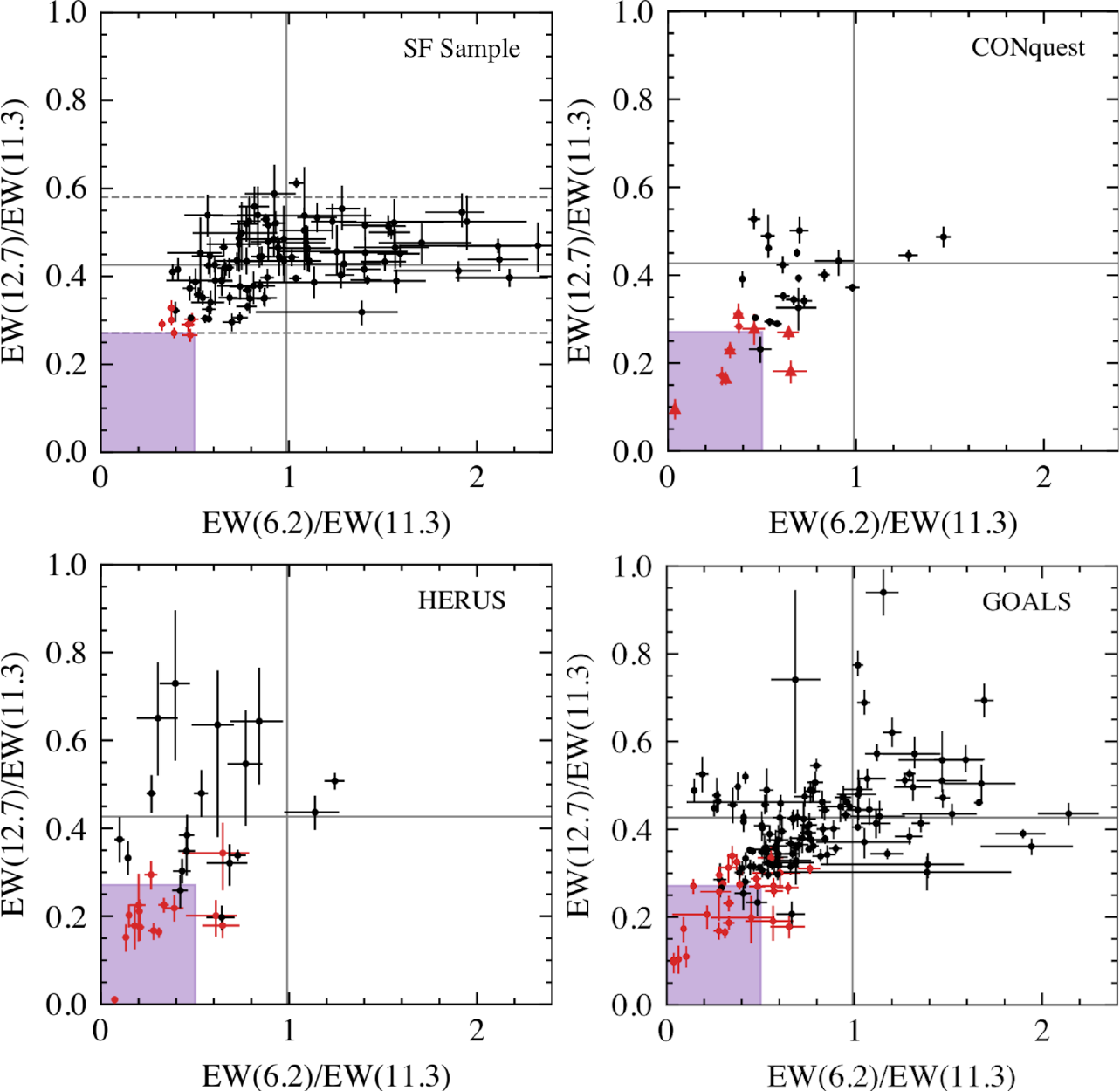}
    \caption{PAH Equivalent Width (EW) diagram \citep[][]{Garcia-Bernete2022} of the 12.7/11.3 PAH features against the 6.2/11.3 PAH features obtained from  fitting the star-forming sample \citep[top left, ][]{Hernan-Caballero2020}, CONquest \citep[top right,][]{Falstad2021}, HERUS \citep[bottom left,][]{Farrah2013}, and GOALS \citep[bottom right,][]{Armus2009}. Points in red have a nuclear optical depth $\tau_N > 3.5$ and for the CONquest sample the triangle points are those identified as CONs by $\Sigma_{\rm HCN-vib} > 1 L_{\odot}$ pc$^{-2}$. The grey lines show the mean values for the star-forming sample with the 2$\sigma$ values for the 12.7/11.3 PAH shown by the dashed lines. The purple shaded area shows the CON selection region.}
    \label{fig:EWRatios} 
\end{figure*}

The resulting EW ratios are shown in the top left panel of Fig. \ref{fig:EWRatios} where the 12.7/11.3 PAH EW ratio has a mean value of $\left< EW_{12.7}/EW_{11.3} \right> = 0.427$ with a scatter of $\sigma \left(EW_{12.7}/EW_{11.3}\right)  = 0.077$. Subtracting the average uncertainty of the data (0.030) in quadrature, we obtain an intrinsic scatter of $\sim 17\%$. The 6.2/11.3 PAH EW ratio shows a much larger scatter with a mean $\left< EW_{6.2}/EW_{11.3} \right> = 0.99$ and a scatter $\sigma \left(EW_{6.2}/EW_{11.3}\right)  = 0.49$.

Comparing our fitting (Drude PAH profiles) to that of IDEOS (Pearson-IV PAH profiles), we find a larger mean value in the $EW_{12.7}/EW_{11.3}$ and a higher scatter, where they found a mean of 0.346 and $\sim 5\%$ intrinsic scatter \citep{Hernan-Caballero2020}. 

The larger scatter in the 6.2/11.3 PAH EW ratio may be due to differences in the intrinsic properties of the continuum. Different dust temperatures will more strongly affect the continuum between 6.2 and 11.3 than 12.7 and 11.3. This can be seen in the left panel of Fig. \ref{fig:SFTemp} where the presence of hotter dust ($> 200$K) will increase the continuum around 6.2 $\mu$m. 

Using the 
newly measured PAH EW values we define the CON selection region as follows: we take the $2\sigma$ lower boundary as the threshold for the 12.7/11.3 and the $1\sigma$ lower limit for the 6.2/11.3. These criteria result in the purple shaded region shown in Fig. \ref{fig:EWRatios}, where the thresholds are $EW_{12.7}/EW_{11.3} = 0.271$ and $EW_{6.2}/EW_{11.3} = 0.503$ for our spectral decomposition tool. 

Two objects in the \citet{Hernan-Caballero2020} star-forming sample fall in the CON region namely VV 283a and UGC 01845. Both of these objects are LIRGs with $\log (L_{\rm IR}/L_{\odot}) = 11.68 $ and $11.12$, respectively \citep[][]{Armus2009}. These two objects are not selected by \citet{Garcia-Bernete2022}. From our optical depth selection discussed in Section \ref{sec:CONDef} we identify a further 5 potential CON candidates from this sample (all of them are LIRGs). Since this sample was 
selected only by excluding AGN, identifying CON candidates
is therefore not unexpected.

\subsection{Physical Properties of CONs}
\label{sec:CONDef}
Fitting the mid-IR spectra with our physically motivated model allows us to recover the continuum of the nuclear component, thus providing insight into the physics of CONs, in particular the nuclear optical depth at 9.8 $\mu$m, $\tau_N$, and the level of dilution from the host galaxy in form of the nuclear fraction, $\beta$. 

\begin{figure*}
 \hspace*{.5cm}                                                           
	\includegraphics[width=17cm]{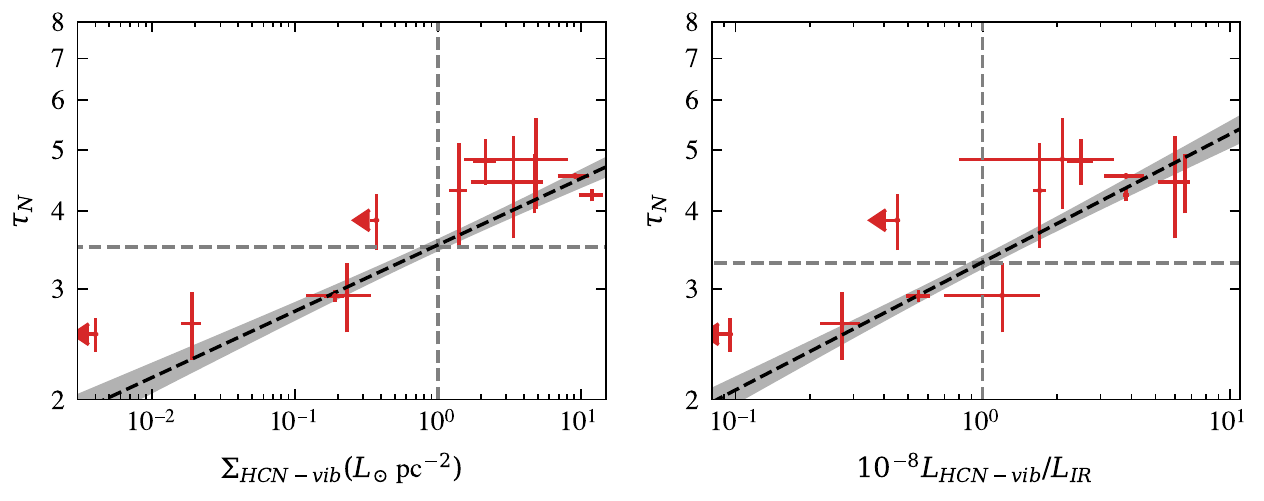}
    \caption{Optical depth of the nuclear component from the spectral decomposition fits to the CONquest sample \citep{Falstad2021} against the surface brightness of HCN-vib (left) and the strength of HCN-vib to $L_{\rm IR}$ (right). Linear fits are shown in the dashed black lines with 1$\sigma$ confidence intervals shown in grey. The upper limits of HCN-vib measurments are shown for objects with $\tau_N>2.5$ and $\beta>0.4$. The grey dashed lines show the CON identification threshold from \citet{Falstad2021} on the x-axis and the corresponding $\tau_N$ on the y-axis.}
    \label{fig:TauPlot} 
\end{figure*}

It is therefore instructive to investigate further properties of the CONs. First, we examine whether there is a possible relation between the model-derived quantities and the properties of the HCN-vib line.
Using our method to fit the CONquest sample from \citet{Falstad2021}, we measure the nuclear optical depth, $\tau_N$. The measured PAH EW ratios are shown in the top right panel of Fig. \ref{fig:EWRatios}. In Fig. \ref{fig:TauPlot} we show $\tau_N$ as a function of the HCN surface brightness in the left panel, and the ratio to the L$_{\rm IR}$ in the right panel. In both plots there is a clear trend for a higher optical depth correlating with stronger HCN-vib emission. A Pearson correlation test results in coefficients of 0.91 and 0.84 for the HCN-vib surface density plot and HCN-vib to $L_{\rm IR}$ plot, respectively. 
This correlation provides good evidence that the heat trapping effect required to populate the vibrational states of HCN is consistent with the presence of large quantities of dust with a high internal temperature producing the mid-IR radiation field and a cool dusty exterior to provide the high optical depths of the silicate absorption feature.

Unlike the nuclear optical depth, the apparent silicate strength as measured using equation \ref{eqn:SilStrength} shows no trend with HCN-vib as this property is highly sensitive to dilution from the host-galaxy. We show this in Appendix \ref{sec:AppE} for reference. We also show the 12.7/11.3 PAH EW ratio against the strength of HCN-vib which does show a trend albeit weaker than the nuclear optical depth.

The \citet{Falstad2021} definition of a CON is based on the strength of the mm vibrational transitions of HCN. They used two definitions, the surface brightness of HCN: $\Sigma_{\rm HCN-vib} > 1 L_{\odot}$ pc$^{-2}$ and the ratio to the IR luminosity: $L_{\rm HCN-vib}/L_{\rm IR} > 10^{-8}$. Based on the correlation found in Fig. \ref{fig:TauPlot} we can therefore attempt to derive a new CON definition based on the optical depth values determined through our fitting method.

To do this we fit a straight line in log-log space using \textsc{scipy} orthogonal distance regression to account for the errors in both x and y. This allows us to find an optical depth corresponding to the definitions of \citet{Falstad2021}. Taking $\Sigma_{\rm HCN-vib} > 1 L_{\odot}$ pc$^{-2}$ requires $\tau_{N} > 3.5$ and $L_{\rm HCN-vib}/L_{\rm IR} > 10^{-8}$ requires $\tau_{N} > 3.3$. As the surface brightness is a more robust definition and gives a stricter threshold, we adopt $\tau_{N} > 3.5$ as a criterion to select deeply obscured nuclei. In addition, we impose the extra condition that the nuclear contribution must be greater than 40$\%$ to ensure the value of $\tau_{N}$ is reliable from the fitting as discussed in Section \ref{sec:Testing}.

This now means we have an additional method to select CON candidates in the mid-infrared based on the decomposition technique presented in this work. 
\subsection{CON Spectral Shape}
\label{sec:SpecShape}
\begin{figure}
\hspace*{-0.3cm}                                                           
	\includegraphics[width=\columnwidth]{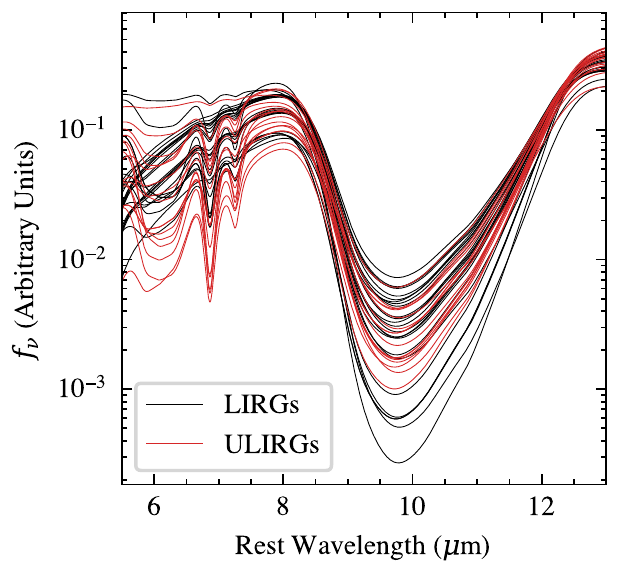}
    \caption{Continua of the nuclear component for LIRGs (black) and ULIRGs (red) selected as CONs from the GOALS and HERUS samples respectively. The spectrum is normalised by the integrated flux between 5 and 15 $\mu$m.}
    \label{fig:NucCont} 
\end{figure}

In Fig. \ref{fig:NucCont} we show the continua of the nuclear components of CONs in LIRGs and ULIRGs from the GOALS and HERUS samples, respectively. These were selected based on the optical depth as described in the following paragraphs. While all show a deep silicate absorption, the slope of the continuum between 5 and 8 $\mu$m varies significantly between objects 
with some displaying flat continua such as ESO 374-IG 032 with others showing steeper spectra, such as the case of IRAS 17578-0400. This may reflect differences in the amount of hot dust present since hotter dust will result in a flatter continuum towards shorter wavelengths. A possible explanation for this might be the presence of 
hotter dust where a more direct line of sight to the central engine allows the dust to reach higher temperatures \citep[e.g.][]{Efstathiou2022, Lyu2018}. This may suggest an inclination dependence where more face-on sources show this hotter dust, however it remains difficult to reconcile this picture with the large levels of obscuration required to produce the deep silicate absorption.

\subsection{Spectral Mapping Sample}

\label{sec:SpecMappingResults}
In this section we present the results of the 
fits 
the spectra from 
from the spectral mapping sample, 
where we fit all the apertures simultaneously (and the shape of the nuclear spectra is the same for all of them, as discussed in Section \ref{sec:Testing}).

\begin{figure*}
\hspace*{1.3cm}                                                           
	\includegraphics[width=16cm]{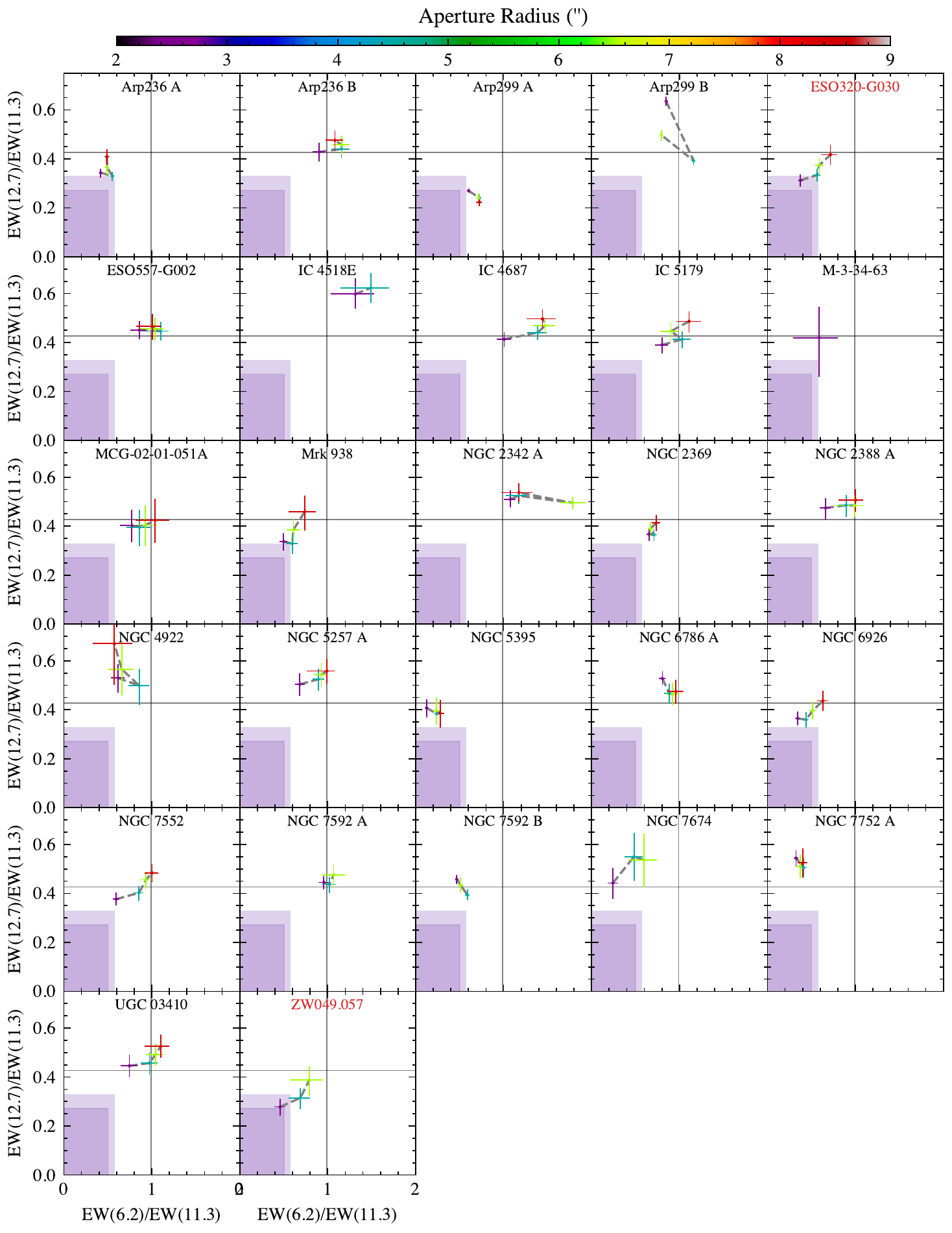}
    \caption{EW ratios of the 12.7/11.3 PAH and the 6.2/11.3 PAH for the spectral mapping sample. For each galaxy, values are colored by the aperture radius used to extract the spectra where a larger aperture will contain more emission from star-formation in the galactic disk. The inner purple box shows the CON selection region as defined for the staring mode spectra, the outer box shows a larger selection region accounting for aperture correction effects. The grey lines show the mean EW ratios for the star-forming calibration sample. The two bona fide CONs are shown with a red title.}
    \label{fig:SpecMapGrid} 
\end{figure*}

Fig. \ref{fig:SpecMapGrid} shows the PAH EW ratios of this sample and how these vary with aperture size. From this figure it is clear that the PAH EW ratio values from the inner aperture deviate from the others for a number of objects e.g. Arp 236 A, Mrk 938. Before analysing further we investigated the effect of the aperture correction on the spectra extracted from this aperture. 
The slope of the continuum is strongly dependent on this correction function and thus the PAH EWs will depend on this. Therefore, due to the aperture corrections, the CON selection box which is calibrated from staring mode spectra may not be directly applicable to spectral maps.

\subsubsection{Aperture Correction Effects}
\label{sec:AppCorr}
To quantify if the aperture correction has a significant effect we compared the PAH EW ratios measured from the spectral mapping mode compared with those from the staring mode. We use NGC 5990 to do this as it has a significant nuclear component

\begin{figure}
	\includegraphics[width=\columnwidth]{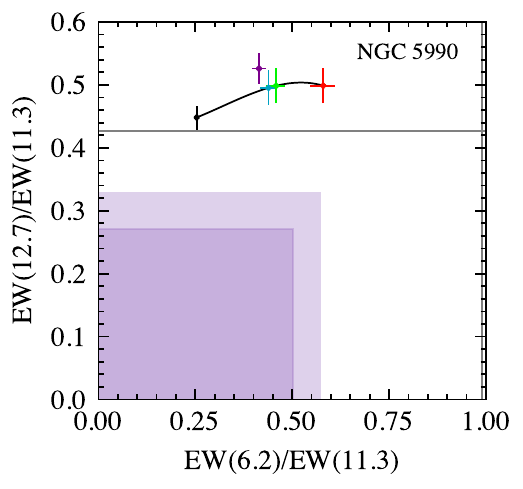}
    \caption{PAH EW ratios for NGC 5990. The coloured points are from spectral mapping observations and are the same as in Fig. \ref{fig:SpecMapGrid}. The additional black point is the staring mode spectra which has a slit radius of approximately 1.8". The purple point has an aperture radius of 2.5" and has an aperture correction applied as described in Section \ref{sec:Obs}. The black line shows a cubic spline interpolation excluding this point. The inner purple box shows the CON selection region as defined for the staring mode spectra, the outer box shows a larger selection region accounting for aperture correction effects. }
    \label{fig:NGC5990} 
\end{figure}

\citep[AGN fraction of 0.49 from][]{Diaz-Santos2017}.  Fig. \ref{fig:NGC5990} shows the PAH EW ratios for each aperture of the spectral maps with the staring mode added as the black point. The aperture of radius 2.5" (in purple) has the aperture correction applied and shows a deviation from the interpolated curve when including the staring mode point but excluding this point. This suggests that the aperture correction may be responsible for changing the PAH EW ratio values for the innermost aperture.

To account for this uncertainty we extend the CON selection criteria to a larger region using the EW ratios for NGC 5990. We use a cubic interpolation of each EW ratio against aperture radius excluding the 2.5'' aperture and measure the deviation of the interpolated vs measured PAH EW ratios at 2.5''. To do this we use a radius of 1.8" for the staring mode spectra as this roughly corresponds to the slit width. Using the measured deviation for the 2.5" aperture we create a larger box shown in Fig. \ref{fig:SpecMapGrid} and \ref{fig:NGC5990} where the 12.7/11.3 EW threshold is increased by 0.058 and the 6.2/11.3 EW threshold is increased by 0.074. This creates an ``error'' region where the EW ratios inferred from the innermost aperture are consistent with a CON.

\subsubsection{CONs in the Spectral Mapping Sample}

Using the adjusted PAH EW CON selection criteria we find two objects that meet both the 12.7/11.3 and the 6.2/11.3 PAH EW criteria: ESO 320-G030 and ZW 049.057, both of which are known CONs with a HCN-vib surface brightness of $\Sigma_{\rm HCN-vib} > 1 L_{\odot}$ pc$^{-2}$ \citep[][]{Falstad2021}.

There are additional objects which may be CONs, selected by only one of the PAH EW criteria. The strongest candidate of these is Arp 299 A as it has a low enough 12.7/11.3 PAH EW ratio which is more reliable than the 6.2/11.3. This object also almost meets the optical depth criterion of  $\tau_N > 3.5$ with $\tau_N = 3.3$. 
Other objects that 
could be CONs are Arp 236 A and Mrk 938 which are both within $1\sigma$ of the adjusted selection region. Additionally, NGC 6926 meets both PAH EW ratios with the staring mode but this spectrum suffers from some reduction issues.

For the two known CONs (ESO 302-G030, ZW049.057) we see a clear trend with how the EW ratios change with aperture size. As more of the galaxy disk is included within the aperture, the EW ratio moves away from the CON selection region and towards the mean values of the star-forming calibration sample. This shows that dilution from the host galaxy can cause objects to be missed by the PAH EW method however higher spatial resolution will overcome this issue. The latter point is important considering the factor $\sim 10$ increase in the spatial resolution of the Mid Infrared Instrument (MIRI) on the James Webb Space Telescope (JWST) compared to the Spitzer spectral maps.

A few other objects show trends with aperture size such as the other potential CON candidates Mrk 938 and NGC 6926 but also NGC 7552 and UGC 03410. Another interesting trend with aperture size can be seen in Arp 299 A, where the 12.7/11.3 PAH EW ratio increases as the aperture decreases.
The high optical depth measured for this object means that the continuum will cause the PAH EW ratio to move in the opposite direction therefore changes in the 12.7/11.3 PAH flux ratio must drive this change and so the PAH properties may be different in the nuclear region.

The flux ratio of PAHs are known to be sensitive to changes in the properties of PAHs such as the size, charge and hardness of the incident radiation field \citep[See][for a review]{Li2020}. There is evidence that the PAH molecules in the vicinity of AGN are larger and more neutral \citep[][]{Garcia-Bernete2022b, Garcia-Bernete2022c} and may even be excited by AGN on very small scales \citep[][]{Jensen2017}. It, therefore, is plausible that Arp 299 A represents an AGN that used to be completely obscured and has expelled some dust exposing PAH molecules to radiation from the central engine and therefore changing the properties of PAH molecules in the nuclear region. Arp 299 B is a known AGN \citep[][]{Alonso-Herrero2009} which also shows a strong increase in the 12.7/11.3 PAH EW ratio with the smallest aperture although this is likely strongly affected by the aperture correction as discussed in Section \ref{sec:AppCorr}.

\section{Discussion}
\label{sec:Discussion}

\subsection{How Many CONs Exist in the Local Universe?}
\label{sec:CONNo}
The first systematic search for CONs was done by \citet{Falstad2021} using the HCN-vib line which found CONs in $38^{+18}_{-13}\%$ of ULIRGs, $21^{+12}_{-6}\%$ of LIRGs and $0^{+9}_{-0}\%$ of sub-LIRGs. This result is limited by the small sample size, hence the large errors. Using the PAH EW technique, \citet{Garcia-Bernete2022} found CONs in $30\%$ of ULIRGs but only $7\%$ of LIRGs, with the discrepancy likely due to dilution of the nuclear continuum emission by star-formation in the disk of the host galaxy. 

To build on this work we can now use the optical depth selection criteria to recover those CONs that were missed by the PAH EW technique to achieve a more accurate estimate of the number of CONs in ULIRGs and LIRGs.

For the ULIRG sample we use HERUS \citep[][]{Farrah2013} as outlined in Section \ref{sec:Obs}. For the LIRG sample we use GOALS \citep[][]{Armus2009}. To estimate 1$\sigma$ uncertainties on the fraction of CONs in each of the sample we follow \citet{Falstad2021} and use \citet{Cameron2011} to construct a beta distribution of the CON fraction which depends on the sample size.

From the HERUS sample of ULIRGs we identify $29^{+7}_{-7}\%$ as CONs from the PAH EW criteria as shown in the bottom left panel of Fig. \ref{fig:EWRatios}, in accordance with \citet{Garcia-Bernete2022}. From the nuclear optical depth definition of $\tau_N > 3.5$ we select slightly more objects with $36^{+8}_{-7}\%$ as CONs. 

The EW ratios of the full GOALS sample are shown in the bottom right panel of Fig. \ref{fig:EWRatios} including the ULIRGs in that sample. From this study, we exclude objects with only spectral maps available as the required aperture corrections may bias the results. Out of the LIRGs in the GOALS sample the EW ratios select $7.7^{+2.3}_{-2.0}\%$ as CONs, consistent with \citet{Garcia-Bernete2022}. The optical depth definition selects $17^{+3}_{-3}\%$ as CONs which is consistent with the CONquest results. In the case of LIRGs, the difference between the optical depth and PAH EW ratios techniques is larger.

\begin{figure}
\hspace*{-0.3cm}                                                           
	\includegraphics[width=\columnwidth]{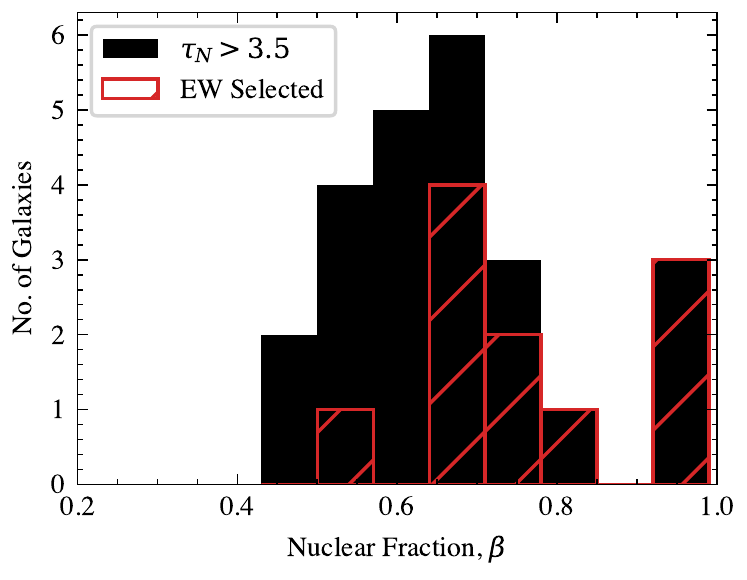}
    \caption{Histogram of the nuclear fraction, $\beta$, for LIRGs in the GOALS sample selected as CONs via the PAH EW method (red) and the nuclear optical depth (black). This shows that the nuclear optical depth identifies additional sources that are more heavily diluted by the host galaxy.}
    \label{fig:NucFracFig} 
\end{figure}

To understand this discrepancy we show in Fig. \ref{fig:NucFracFig} the measured nuclear fraction, $\beta$, for the LIRGs selected as hosting CONs with each technique. The plot shows that the nuclear optical depth method selects additional sources with lower nuclear fractions and thus greater dilution from the host galaxy, which the PAH EW method misses. The discrepancy being larger for LIRGs compared to ULIRGs 
is a consequence of the fact that LIRGs contain more extended star-forming components and thus are more susceptible to dilution from the host galaxy. This is consistent with what was observed with the spectral mapping data in Section \ref{sec:SpecMappingResults}.

For both LIRGs and ULIRGs the fraction of CONs is consistent with the results of the CONquest investigation \citep[][]{Falstad2021} with overlapping 1$\sigma$ intervals. A summary comparing the various methods is given in Table \ref{tab:CONNos}. The larger sample sizes afforded by using mid-IR observations results in tighter constraints on these numbers. It is worth noting that the peak of the distribution for the number of CONs in LIRGs is slightly lower than the CONquest results. This may just be statistical error but it could also be due to objects where the nucleus is so diluted by the host galaxy that they are excluded by our cut requiring a nuclear contribution of 40\%. 

\begin{table}
\centering
  \caption{Fraction of CONs in ULIRGs and LIRGs in the local Universe}
  \label{tab:CONNos}
    \def\arraystretch{1.2}
    \setlength{\tabcolsep}{4pt}
    \begin{threeparttable}
  \begin{tabular}{ccc}
  
    \hline

     Method & ULIRGs & LIRGs \\
    \hline
    HCN-vib \citep[][]{Falstad2021} &$38^{+18}_{-13}\%$ & $21^{+12}_{-6}\%$\\
    PAH EW \citep[][]{Garcia-Bernete2022} & $30\%$& $7.0\%$ \\
    PAH EW (This Work) & $29^{+7}_{-7}\%$ &  $7.7^{+2.3}_{-2.0}\%$\\
    Nuclear Optical Depth, $\tau_N$ (This Work) & $36^{+8}_{-7}\%$ & $17^{+3}_{-3}\%$\\

    \hline
  
  \end{tabular}

  \end{threeparttable}
 \end{table}

A comprehensive table of all the galaxies analysed in this work can be found in Appendix \ref{sec:AppC}, where measured PAH EW ratios and nuclear fraction/optical depths are reported. We also report on the detection of the HCN 14 $\mu$m absorption line \citep{Lahuis2007}.
Finally, we also include information on the detections 
of the 23 $\mu$m, 28 $\mu$m or 33 $\mu$m crystalline silicate absorption features 
from \citet{IDEOS}. These features in absorption indicate heavily obscured nuclei via two methods. Method I indicates sources with the 23 $\mu$m feature and $33\mu$m feature in absorption i.e. $s_{23}<0.0$ and $s_{33}<0$. Method II indicates sources with $s_{23}<-0.09$ and $s_{23}<-0.02$. \citet{IDEOS} also includes a method III for detection of the blue wing of the 33 $\mu$m feature where the full feature is cut due to the redshift of the source. No objects analysed in this work are classified by method III.

\subsection{Effect of Galaxy Inclination}
\label{sec:Inclination}
As mentioned previously the PAH EW method can select CON candidates independent of galaxy inclination as extinction from the star-forming disk will affect both the PAH flux ratio and the continuum ratio leaving the EW ratio unchanged. However this is not necessarily the case for the nuclear optical depth selection. In Fig. \ref{fig:EWRatios} there is one object, NGC 3628, with a $\tau_N > 3.5$ but is not selected as a CON by either the PAH EW ratios or the HCN-vib surface brightness. This is a highly inclined galaxy which suggests that the nuclear optical depth obtained from our spectral decomposition fitting may be sensitive to such galaxies with strong dust lanes obscuring the line of sight, which may bias our estimates of the number of CONs.

\begin{figure}
	\includegraphics[width=\columnwidth]{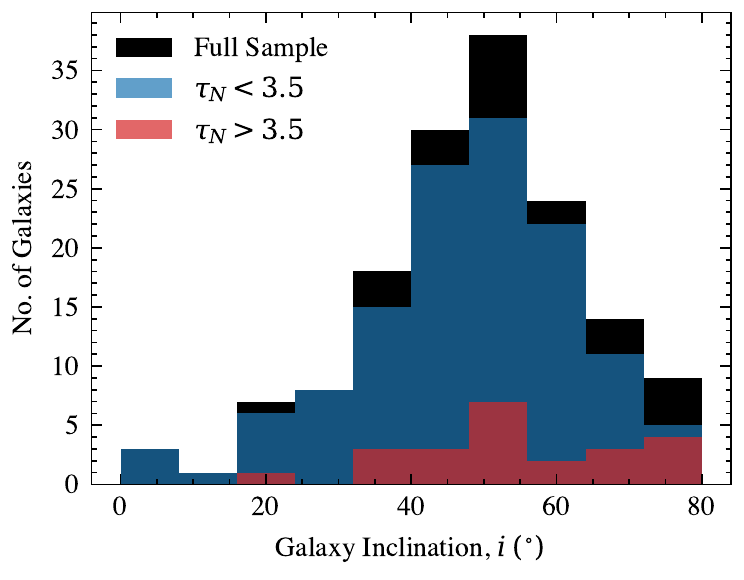}
    \caption{Inclination of galaxies in the GOALS sample obtained from axis ratios. The full sample is shown in black with objects selected as obscured nuclei from the optical depth shown in red and those not as CONs in blue.}
    \label{fig:InclinationFig} 
\end{figure}

To quantify if this is an issue we obtained estimates for the inclination of galaxies in the GOALS sample from the NASA/IPAC Extragalactic Database (NED)\footnote{\url{https://ned.ipac.caltech.edu/}} using the ratio of the minor axis, $b$, to the major axis, $a$, to estimate the inclination, $i$ where $\cos i = b/a$. For the majority of the sample, the axis ratios were from 2MASS imaging in the $\rm K_s$ band \citep[][]{2MASS} or r band imaging from SDSS \citep[][]{SDSS}. We select a sub-sample with $\tau_N>3.5$ and compare to the full sample. This is shown in Fig. \ref{fig:InclinationFig}.

Visually the two samples look similar however there may be some extra galaxies selected at the highest inclinations ($i \gtrsim 65^{\circ}$). To determine if there is any statistically significant difference between the samples we perform a two-sample Kolmogorov-Smirnov (K-S) test \citep{KSTest} from the {\sc scipy.stats} package. This tests the null-hypothesis that the two samples are drawn from the same underlying distribution which typically requires a p-value $< 0.05$ to reject. We perform the K-S test on the sample with $\tau_N <3.5$ and $\tau_N>3.5$ to test whether there is any significant difference. We find a p-value of $0.43$ which suggests there is no statistically significant difference between the two samples. It is worth noting however that the small sample size of the CONs may make the K-S test less likely to find a difference even if one exists and so it is worth inspecting visually. We would also only expect a difference at high inclinations where the dust lanes would obscure the line of sight towards the nucleus and so if we only compare the highest inclinations ($i \gtrsim 65^{\circ}$), we do see a bias where sources may be falsely selected as CON candidates. The case of NGC 3628 is one such example however this case also shows the strength of the PAH EW method to select CON candidates.

While there may be a bias falsely selecting highly inclined sources as CON candidates, this bias may actually counteract the exclusion of the most diluted CONs by their host galaxy ($\beta < 0.4$) and so while individual objects may be miss-classified, the total number of CONs likely remains accurate.

\subsection{Future Prospects}
To really test the impact of CONs on galaxy evolution requires observations across cosmic time, particularly at cosmic-noon ($z \sim 1-3$). The advent of the James Webb Space Telescope (JWST) will be at the forefront of finding and understanding CONs beyond the local universe \citep[][]{Garcia-Bernete2022}. In particular the Mid-InfraRed Instrument for JWST \citep[MIRI, ][]{Rieke2015}, will enable imaging \citep{Wright2015} and spectroscopy \citep{Wells2015} of galaxies between 5 and 28 $\mu$m. 

We found in this work that dilution of the nuclear emission by relatively unobscured star-formation is a challenge for identifying CONs with Spitzer. As the JWST provides a nearly 10 times increase in angular resolution, this issue will become apparent at $z \sim 0.5$ where the physical scale probed will surpass that of local galaxies with Spitzer.

Future surveys in the mid-infrared are also important as these will enable analysis of large numbers of these objects at cosmic noon such as PRIMA \citep[][]{GEP}.

\section{Conclusions}
In this work we have investigated the physical properties of compact obscured nuclei by decomposing the mid-IR spectra into nuclear and star-forming and evaluated how to identify such objects using PAH EW ratios. We have made the code is publicly available \footnote{\url{https://github.com/FergusDonnan/PAHDecomp}}. Our main findings are:
\begin{itemize}

    \item From our spectral decomposition, the optical depth of the nuclear component at 9.8 $\mu$m, $\tau_N$, is strongly correlated with the surface brightness/strength of HCN-vib emission at millimeter wavelengths, suggesting the same physics is responsible for both observational signatures. This leads to a CON selection criteria of $\tau_N > 3.5$.
    
    \item From $\tau_N > 3.5$ we find that CONs make up $36^{+8}_{-7}\%$ of ULIRGs and $17^{+3}_{-3}\%$ of LIRGs, consistent with the results of CONquest but with tighter constraints. We find the PAH EW method classifies fewer CONs likely due to the low spatial resolution of Spitzer IRS data where this method detects $29^{+7}_{-7}\%$ of ULRIGs and $7.7^{+2.3}_{-2.0}\%$ of LIRGs as CONs.

    \item We find that the PAH EW method is robust against false positives in highly inclined galaxies with strong dust lanes that can produce high optical depths, whereas using the nuclear optical depth to select CON sources may falsely select some objects that are highly inclined.
    
    \item From spectra extracted at different spatial scales we find the PAH EW ratios move towards the CON selection criteria with smaller apertures where there is less dilution from the disk. This suggests that star-formation diluting the nucleus is responsible for the EW ratios underestimating the number of LIRGs hosting CONs.
\end{itemize}
We have confirmed that the mid-infrared can be used to effectively to select completely obscured nuclei and investigate properties of the dust. This will allow these objects to be found/studied beyond the local universe where HCN-vib emission is simply too faint to be detected, even with ALMA. With its high sensitivity, the JWST will allow us to find and understand these objects at cosmic noon where the impact of this powerful but hidden phase of galaxy evolution can be uncovered.

\begin{acknowledgements}
      FRD acknowledges support from STFC through grant ST/W507726/1. DR and IGB acknowledge support from STFC through grant ST/S000488/1. DR also acknowledges support from the University of Oxford John Fell Fund. AAH acknowledges support from grant PGC2018-094671-B-I00 funded by MCIN/AEI/ 10.13039/501100011033 and by ERDF A way of making Europe, and grant PID2021-124665NB-I00. We thank Michalina Maksymowicz-Maclata for providing the list crystalline absorption detections. We also thank the reviewer for the useful feedback.

\end{acknowledgements}

%
   \bibliographystyle{aa} 
   \bibliography{References} 
%

\appendix
\section{Extinction Templates}
\label{sec:AppA}
There are absorption features observed between 5.5 - 8 $\mu$m in strongly obscured galaxies, commonly attributed to water ice absorption at $\sim 6 \mu$m \citep{Spoon2001} and the deformation mode of aliphatic CH molecules at $6.85$ and $7.25 \mu$m \citep{Dartois2007} in addition to the deep silicate absorption at $\sim 9.8\mu$m. A theoretical template remains elusive as the molecular composition responsible for this absorption is very complex, therefore we follow \citet{IDEOS} and resort to constructing a template from a heavily obscured source that has very little PAH emission contaminating the region, namely NGC 4418. 

\begin{figure*}
\hspace*{0.5cm}                                                           
	\includegraphics[width=16cm]{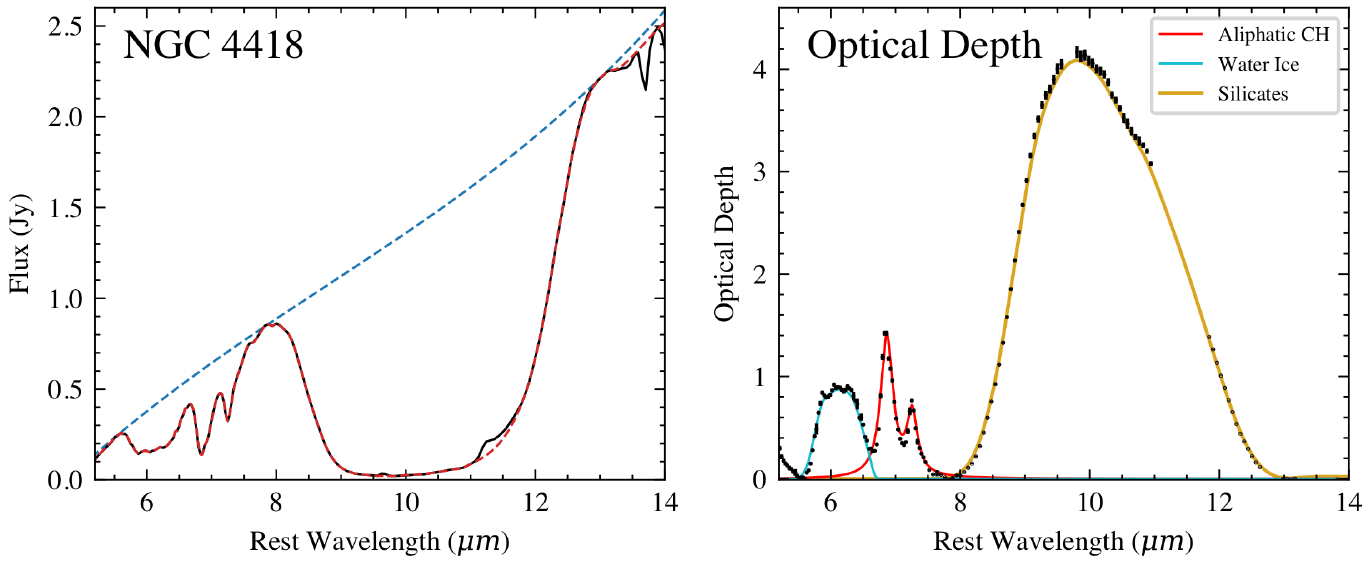}
   \caption{{\it Left:} IRS spectrum of NGC 4418 (black) with the spline interpolated underlying continuum in the dashed blue line. The dashed red line shows the spectrum with emission features removed. {\it Right:} Optical depth profile using the underlying continuum, overlayed with the inferred template for water ice (cyan), aliphatic CH (red) absorption and silicates (gold).}
    \label{fig:ExtTemplate} 
\end{figure*}

Using Spitzer IRS spectra of NGC 4418, we determine the local underlying continuum for through a cubic spline interpolation anchored at 5.5, 7.8, 13.0 and 14.5 $\mu$m. This is shown in the left panel of Fig. \ref{fig:ExtTemplate} as a blue dashed line. To extract an optical depth, we first mask out any PAH emission or lines before taking the natural log of the ratio of the spline continuum to the masked data. This results in the optical depth in the right panel of Fig. \ref{fig:ExtTemplate}.

We fit the optical depth profile with three components, the ice, CH and silicates. \citet{IDEOS} used two Gaussian profiles centred at $6.85$ and $7.25 \mu$m to represent the CH deformation mode however we find the fit is poor, requiring broader wings to fit the optical depth spectrum. Therefore we fit two Drude profiles to the optical depth spectrum instead, which is shown in red in the right panel of Fig. \ref{fig:ExtTemplate} (Lorentzian profiles also provided a good fit however the $\chi^2$ was slightly higher). This provides the template for the CH absorption, while the ice feature is given by the residuals smoothed to provide a template, shown in cyan in the figure. The silicate profile is also given by smoothing the residuals.

\section{Choice of Silicate Template}
\label{sec:AppD}
In our analysis we chose to use an empirical template for the 9.8 $\mu$m silicate absorption feature, derived from NGC 4418 as described in Appendix \ref{sec:AppA}. We chose this source as it is highly obscured with minimal emission features and is a well studied CON with strong HCN-vib emission \citep[e.g.][]{Sakamoto2010}. However as noted in Section \ref{sec:NucComp}, ground based observations with a smaller beam find a higher peak optical depth than with Spitzer IRS which suggests the template contains some contribution from the relatively unobscured continuum from circumnuclear star-formation. We therefore test another template derived from another highly obscured galaxy, IRAS 08572+3915. 

\begin{figure*}
\hspace*{0.5cm}                                                           
	\includegraphics[width=17cm]{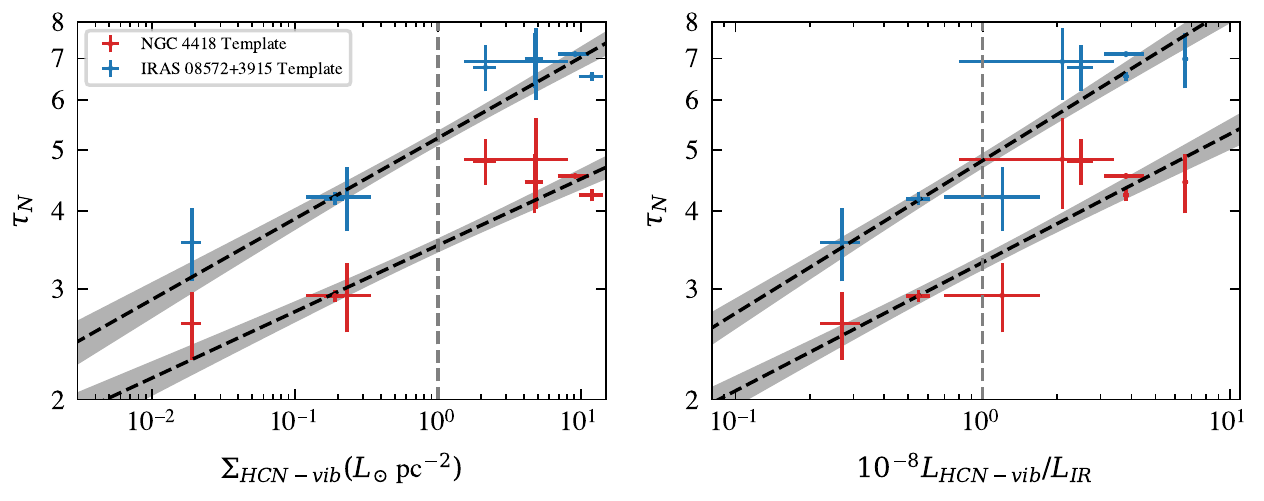}
    \caption{Comparison of the measured nuclear optical depths of the CONquest sample with different silicate templates. In red are the measured $\tau_N$ against the surface density of HCN-vib (left) and HCN-vib luminosity to $L_{\rm IR}$ (right) using the NGC 4418 profile. In blue are the results using the template from IRAS 08572+3915.}  
    \label{fig:TauFigApp} 
\end{figure*}

The profile for this galaxy is narrower which suggests less contamination by any unobscured continuum, consistent with a total absence of emission features. We tested the model using this template on the CONquest sample and compared the resulting decomposition. In Fig. \ref{fig:TauFigApp} we compare the measured optical depths using the IRAS 08572+3915 template in blue to the NGC 4418 template in red. As expected the IRAS 08572 template produces higher nuclear optical depths requiring more contribution from the star-forming component to fit the data. Crucially this is a constant offset in optical depth with the same correlation found but shifted to higher values. Therefore the conclusions of this paper are not strongly affected by the choice of silicate template.

While this template may be advantageous for future work, we chose not to use it as this object is not a CON by the HCN-vib definition.

\section{Silicate Strength vs HCN-vib}
\label{sec:AppE}
\begin{figure*}
\hspace*{0.5cm}                                                           
	\includegraphics[width=17cm]{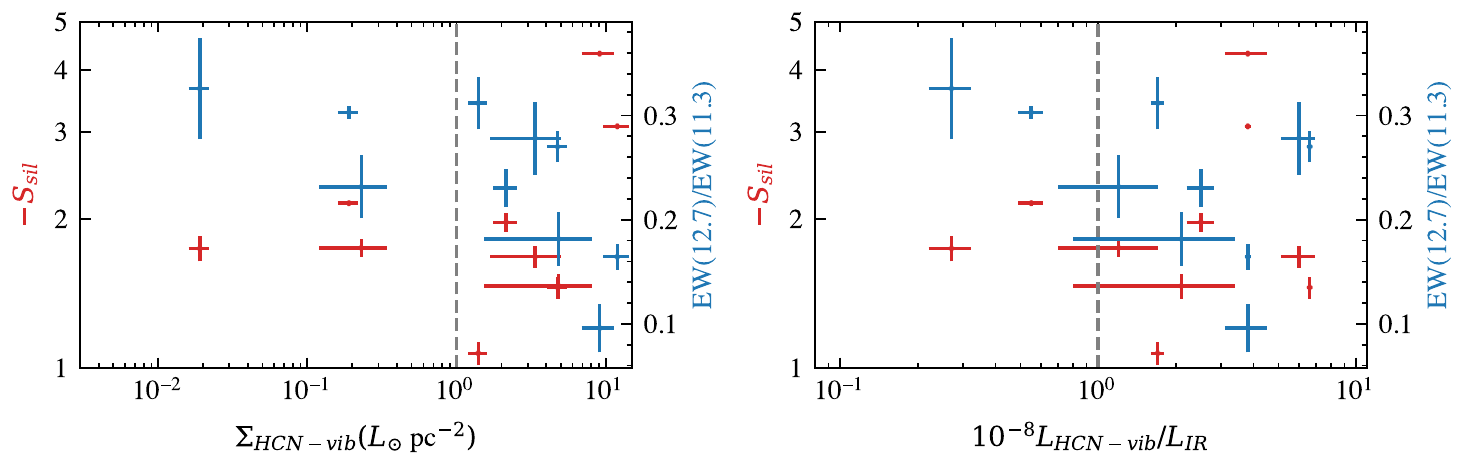}
    \caption{Same as Fig. \ref{fig:TauPlot} but with the apparent silicate strength as calculated by equation \ref{eqn:SilStrength} (red) which shows no trend. The 12.7/11.3 PAH EW ratio is also shown (blue) against HCN-vib on the right hand axis of each plot and shows a very weak trend. Comparing to Fig. \ref{fig:TauPlot} shows the value in accounting for the star-forming contribution to properly recover properties of the obscured nucleus. }
    \label{fig:HCNTrend} 
\end{figure*}

\section{Continuum Ratios of Spectral Mapping Sample}
\label{sec:AppB}
\begin{figure*}
\hspace*{0.5cm}                                                           
	\includegraphics[width=17cm]{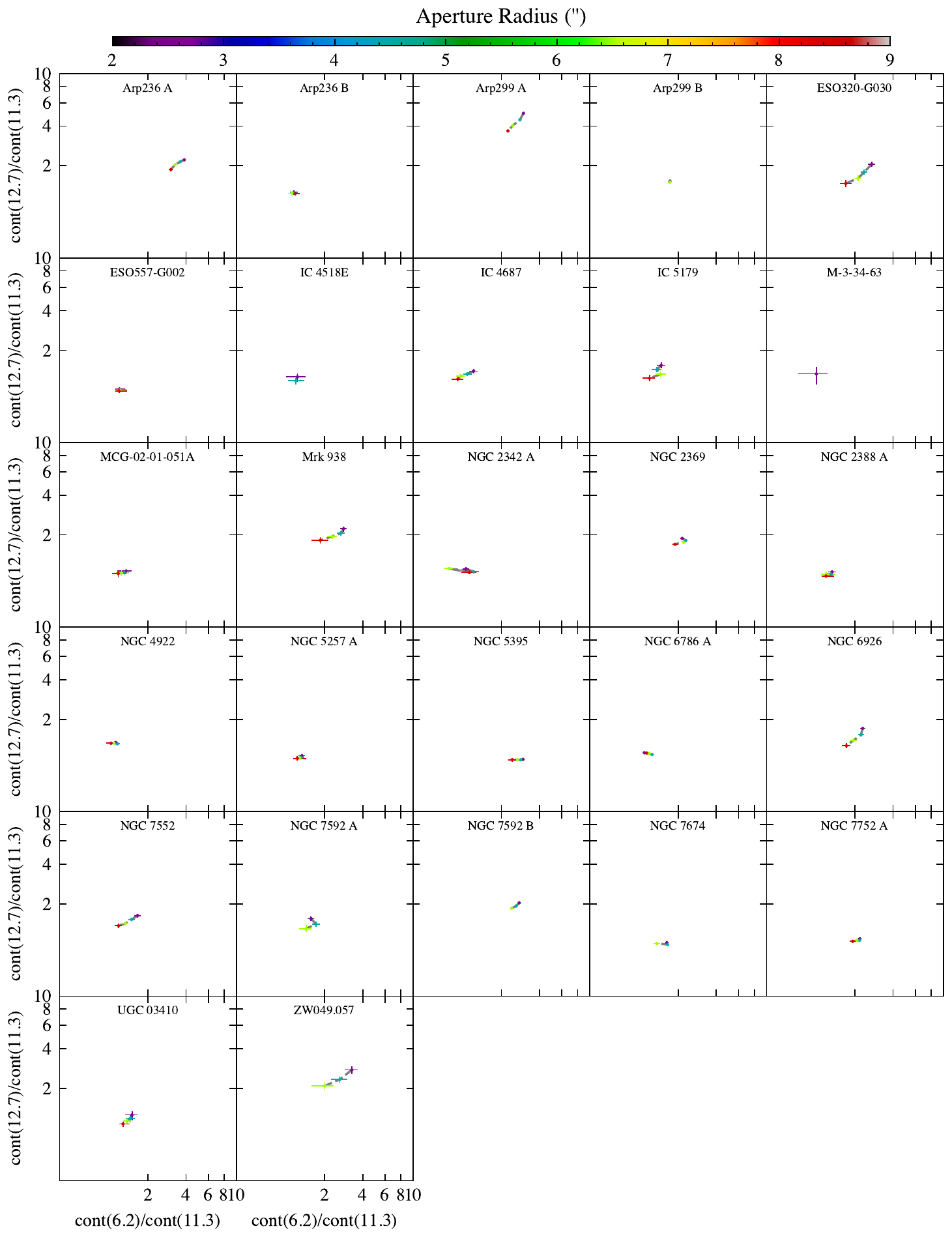}
    \caption{Continuum ratios at 12.7 $\mu$m to 11.3 $\mu$m against the 6.2/11.3 continuum ratio for the spectral mapping sample. For each galaxy, values are colored by the aperture radius used to extract the spectra where a larger aperture will contain more emission from star-formation in the galactic disk.}
    \label{fig:ContRatioGrid} 
\end{figure*}

\section{Table of Spectral Properties}
\label{sec:AppC}
\begin{table*}
\centering
  \caption{Spectral Properties of all the Galaxies used in this Work}
  \label{tab:Properties}
    \def\arraystretch{1.2}
    \setlength{\tabcolsep}{1pt}
    \begin{threeparttable}
   \begin{tabular}{cccccccccc}
  
     \hline

    Name & EW(6.2)/EW(11.3) & EW(12.7)/EW(11.3) & $S_{\rm Sil}$ & $\tau_N$ & $\beta$ & HCN (14 $\mu$m) & Crystallines?\\
    (1) & (2) & (3) & (4) & (5) & (6) & (7) & (8)\\
    \hline
2MASS J03574895-1340458& $0.923^{+0.114}_{-0.157}$ & $0.588^{+0.065}_{-0.068}$ & $-0.462^{+0.083}_{-0.104}$ & - & - &  -  &-\\
2MASX J00480675-2848187& $0.792^{+0.083}_{-0.103}$ & $0.350^{+0.031}_{-0.034}$ & $-1.061^{+0.092}_{-0.100}$ & - & -&  -  &-\\
2MASX J02245768-0414182& $1.082^{+0.356}_{-0.486}$ & $0.538^{+0.111}_{-0.124}$ & $-0.661^{+0.221}_{-0.146}$ & - & -&  -  &-\\
2MASX J02253645-0500123& $1.081^{+0.15}_{-0.129}$ & $0.505^{+0.052}_{-0.047}$ & $-0.433^{+0.106}_{-0.099}$ & - & -&  -  &-\\
2MASX J05583717-7716393& $0.776^{+0.075}_{-0.085}$ & $0.434^{+0.032}_{-0.029}$ & $-1.402^{+0.079}_{-0.08}$ & - & - &  - &-\\
2MASX J08182925+3717481& $0.973^{+0.226}_{-0.256}$ & $0.485^{+0.076}_{-0.083}$ & $-0.641^{+0.217}_{-0.096}$ & - & - &  - &-\\
2MASX J09192731+3347270& $0.974^{+0.196}_{-0.37}$ & $0.435^{+0.040}_{-0.049}$ & $-0.725^{+0.182}_{-0.304}$ & - & - &  - &-\\
2MASX J10363621+6322224& $2.325^{+0.486}_{-0.742}$ & $0.470^{+0.053}_{-0.061}$ & $-0.444^{+0.117}_{-0.096}$ & - & -&  -  &-\\
2MASX J11182408+5602074& $1.947^{+0.319}_{-0.418}$ & $0.525^{+0.06}_{-0.065}$ & $-0.445^{+0.112}_{-0.099}$ & - & - &  - &-\\
2MASX J14081899+2904474& $1.56^{+0.266}_{-0.292}$ & $0.523^{+0.053}_{-0.058}$ & $-0.542^{+0.172}_{-0.153}$ & - & - &  - &-\\
2MASX J14094683-0820036& $1.566^{+0.18}_{-0.224}$ & $0.466^{+0.043}_{-0.048}$ & $-0.428^{+0.103}_{-0.089}$ & - & - &  - &-\\
2MASX J14255448+3446026& $1.405^{+0.13}_{-0.125}$ & $0.516^{+0.041}_{-0.035}$ & $-0.192^{+0.08}_{-0.072}$ & - & - &  - &-\\
2MASX J14520570+3810593& $1.091^{+0.102}_{-0.11}$ & $0.477^{+0.040}_{-0.042}$ & $-0.221^{+0.087}_{-0.076}$ & - & -&  -  &-\\
2MASX J15574349+2727530& $1.134^{+0.294}_{-0.326}$ & $0.386^{+0.036}_{-0.038}$ & $-0.789^{+0.206}_{-0.104}$ & - & - &  - &-\\
2MASX J16070059+5538090& $1.706^{+0.265}_{-0.229}$ & $0.477^{+0.043}_{-0.047}$ & $-0.129^{+0.056}_{-0.040}$ & - & -&  -  &-\\
2MASX J16140266+5330358& $1.15^{+0.105}_{-0.106}$ & $0.534^{+0.032}_{-0.034}$ & $-0.069^{+0.046}_{-0.026}$ & - & - &  - &-\\
2MASX J16164521+5502305& $0.735^{+0.119}_{-0.173}$ & $0.473^{+0.051}_{-0.063}$ & $-0.589^{+0.146}_{-0.183}$ & - & - & \checkmark &-\\
2MASX J18003399-0401443& $1.283^{+0.094}_{-0.088}$ & $0.554^{+0.053}_{-0.050}$ & $-0.14^{+0.077}_{-0.061}$ & - & - &  - &-\\
2MASX J18113842+0131397& $0.78^{+0.050}_{-0.060}$ & $0.332^{+0.009}_{-0.010}$ & $-0.861^{+0.037}_{-0.039}$ & - & - &  - &-\\
2MASX J18324117-3411274& $0.575^{+0.015}_{-0.017}$ & $0.303^{+0.007}_{-0.008}$ & $-1.531^{+0.020}_{-0.022}$ & - & - &  - &-\\
2MASX J19565118+1633389& $0.608^{+0.090}_{-0.262}$ & $0.391^{+0.040}_{-0.058}$ & $-0.752^{+0.111}_{-0.247}$ & - & - &  - &-\\
2MASX J21270303+2355456& $0.774^{+0.065}_{-0.082}$ & $0.519^{+0.039}_{-0.040}$ & $-0.396^{+0.087}_{-0.084}$ & - & - &  - &-\\
2MASX J22382548-1646485& $1.232^{+0.128}_{-0.115}$ & $0.525^{+0.04}_{-0.041}$ & $-0.173^{+0.098}_{-0.057}$ & - & - &  - &-\\
2MFGC 13321& $0.529^{+0.114}_{-0.177}$ & $0.453^{+0.081}_{-0.087}$ & $-0.869^{+0.230}_{-0.183}$ & - & - &  - &-\\
3C 273& $\mathbf{0.396^{+0.078}_{-0.084}}$ & $0.730^{+0.166}_{-0.176}$ & $0.126^{+0.011}_{-0.011}$ & - & - &  - &-\\
AM 0702-601 NED02& $0.891^{+0.030}_{-0.033}$ & $0.517^{+0.025}_{-0.025}$ & $-0.686^{+0.033}_{-0.034}$ & - & - &  - &-\\
Arp 148& $0.509^{+0.034}_{-0.036}$ & $0.307^{+0.014}_{-0.017}$ & $-1.379^{+0.037}_{-0.043}$ & $3.430^{+0.621}_{-0.442}$ & $0.589^{+0.032}_{-0.030}$&  - &-\\
Arp 220& $\mathbf{0.308^{+0.016}_{-0.016}}$ & $\mathbf{0.164^{+0.010}_{-0.014}}$ & $-3.078^{+0.020}_{-0.020}$ & $\mathbf{4.237^{+0.098}_{-0.094}}$ & $0.89^{+0.006}_{-0.006}$ & \checkmark & I \\
Arp 256 NED01& $1.401^{+0.086}_{-0.098}$ & $0.416^{+0.016}_{-0.02}$ & $-0.624^{+0.037}_{-0.036}$ & - & - &  - &-\\
Arp 295 B& $0.739^{+0.043}_{-0.048}$ & $0.306^{+0.013}_{-0.014}$ & $-0.919^{+0.04}_{-0.044}$ & - & - &  - &-\\
Arp236 A$^*$& $\mathbf{0.418^{+0.011}_{-0.012}}$ & $0.343^{+0.017}_{-0.02}$ & $-1.611^{+0.017}_{-0.019}$ & - & -&  -  &-\\
Arp236 B$^*$& $0.904^{+0.065}_{-0.073}$ & $0.428^{+0.039}_{-0.038}$ & $-0.549^{+0.061}_{-0.062}$ & - & - &  - &-\\
Arp299 A$^*$& $0.603^{+0.02}_{-0.02}$ & $\mathbf{0.269^{+0.009}_{-0.009}}$ & $-2.412^{+0.026}_{-0.027}$ & $3.256^{+0.049}_{-0.047}$ & $0.881^{+0.007}_{-0.007}$&  - &-\\
Arp299 B$^*$& $0.849^{+0.015}_{-0.016}$ & $0.635^{+0.018}_{-0.018}$ & $-0.874^{+0.003}_{-0.003}$ & - & - &  - &-\\
CGCG 011-076& $\mathbf{0.446^{+0.019}_{-0.021}}$ & $0.315^{+0.016}_{-0.02}$ & $-1.031^{+0.038}_{-0.04}$ & - & - &  - &-\\
CGCG 052-037& $0.782^{+0.029}_{-0.034}$ & $0.362^{+0.012}_{-0.016}$ & $-0.879^{+0.027}_{-0.030}$ & - & -&  -  &-\\
CGCG 058-009& $\mathbf{0.142^{+0.007}_{-0.007}}$ & $\mathbf{0.271^{+0.016}_{-0.018}}$ & $-1.798^{+0.049}_{-0.048}$ & $\mathbf{4.905^{+0.389}_{-0.329}}$ & $0.653^{+0.023}_{-0.025}$&  - &-\\
CGCG 141-034& $\mathbf{0.403^{+0.024}_{-0.027}}$ & $0.312^{+0.016}_{-0.018}$ & $-1.626^{+0.064}_{-0.061}$ & $3.004^{+0.219}_{-0.203}$ & $0.638^{+0.036}_{-0.039}$&  - &-\\
CGCG 142-034& $\mathbf{0.349^{+0.027}_{-0.028}}$ & $0.340^{+0.023}_{-0.023}$ & $-1.401^{+0.061}_{-0.054}$ & $\mathbf{3.750^{+0.762}_{-0.507}}$ & $0.491^{+0.056}_{-0.048}$&  - &-\\
CGCG 152-070& $0.812^{+0.114}_{-0.193}$ & $0.378^{+0.032}_{-0.041}$ & $-0.507^{+0.114}_{-0.188}$ & - & - &  - &-\\
CGCG 290-067& $0.835^{+0.091}_{-0.113}$ & $0.539^{+0.065}_{-0.072}$ & $-0.515^{+0.088}_{-0.082}$ & - & - &  - &-\\
CGCG 436-030& $0.660^{+0.045}_{-0.048}$ & $0.315^{+0.018}_{-0.019}$ & $-1.588^{+0.047}_{-0.049}$ & - & - &  - &-\\
CGCG 453-062& $0.592^{+0.054}_{-0.067}$ & $0.337^{+0.025}_{-0.031}$ & $-1.043^{+0.086}_{-0.099}$ & $2.654^{+0.967}_{-0.534}$ & $0.476^{+0.071}_{-0.071}$&  - &-\\
CGCG 465-012& $1.120^{+0.312}_{-0.064}$ & $0.572^{+0.022}_{-0.017}$ & $-0.101^{+0.32}_{-0.019}$ & - & - &  - &-\\
CGCG 468-002 NED02& $0.872^{+0.064}_{-0.075}$ & $0.351^{+0.024}_{-0.020}$ & $-1.294^{+0.049}_{-0.050}$ & - & - &  - &-\\
CGCG 468-002& $\mathbf{0.353^{+0.025}_{-0.028}}$ & $0.456^{+0.039}_{-0.042}$ & $-0.121^{+0.019}_{-0.019}$ & - & - &  - &-\\
CXO J191431.2-211905& $0.576^{+0.036}_{-0.039}$ & $0.325^{+0.022}_{-0.023}$ & $-1.316^{+0.046}_{-0.056}$ & $2.785^{+0.269}_{-0.253}$ & $0.64^{+0.034}_{-0.039}$&  - &-\\
     \hline
  
   \end{tabular}
   \end{threeparttable}
  \end{table*}

 \begin{table*}
 \ContinuedFloat  
 \centering
   \caption{Spectral Properties Cont.}
     \def\arraystretch{1.2}
    \setlength{\tabcolsep}{1pt}
    \begin{threeparttable}
   \begin{tabular}{cccccccccc}
  
     \hline

    Name & EW(6.2)/EW(11.3) & EW(12.7)/EW(11.3) & $S_{\rm Sil}$ & $\tau_N$ & $\beta$ & HCN (14 $\mu$m) & Crystallines?\\
    (1) & (2) & (3) & (4) & (5) & (6) & (7) & (8)\\
     \hline
     
ESO 069-IG 006N& $1.269^{+0.042}_{-0.050}$ & $0.512^{+0.015}_{-0.015}$ & $-0.661^{+0.018}_{-0.013}$ & - & - &  - &-\\
ESO 099-G004& $1.176^{+0.085}_{-0.090}$ & $0.344^{+0.011}_{-0.012}$ & $-0.953^{+0.059}_{-0.052}$ & - & - &  - &-\\
ESO 138-G027& $1.354^{+0.048}_{-0.049}$ & $0.414^{+0.020}_{-0.020}$ & $-0.267^{+0.023}_{-0.024}$ & - & - &  - &-\\
ESO 148-IG 002& $1.244^{+0.052}_{-0.055}$ & $0.508^{+0.018}_{-0.02}$ & $-0.698^{+0.042}_{-0.045}$ & - & - &  - &-\\
ESO 173-G015& $\mathbf{0.496^{+0.015}_{-0.016}}$ & $0.314^{+0.006}_{-0.006}$ & $-2.144^{+0.020}_{-0.020}$ & $2.887^{+0.071}_{-0.066}$ & $0.813^{+0.011}_{-0.011}$&  - &-\\
ESO 221-IG10& $0.908^{+0.081}_{-0.126}$ & $0.432^{+0.026}_{-0.034}$ & $-0.441^{+0.053}_{-0.083}$ & - & - &  - &-\\
ESO 239-IG002& $1.676^{+0.182}_{-0.199}$ & $0.504^{+0.043}_{-0.043}$ & $-0.442^{+0.076}_{-0.065}$ & - & - &  - &-\\
ESO 244- G 012 NED02& $1.511^{+0.112}_{-0.129}$ & $0.433^{+0.021}_{-0.022}$ & $-0.767^{+0.044}_{-0.044}$ & - & - &  - &-\\
ESO 255-IG 007 NED01& $2.113^{+0.177}_{-0.220}$ & $0.469^{+0.017}_{-0.016}$ & $-0.300^{+0.037}_{-0.025}$ & - & - &  - &-\\
ESO 255-IG 007 NED03& $1.406^{+0.217}_{-0.244}$ & $0.454^{+0.034}_{-0.036}$ & $-0.366^{+0.099}_{-0.082}$ & - & - &  - &-\\
ESO 255-IG007& $1.593^{+0.101}_{-0.167}$ & $0.559^{+0.033}_{-0.029}$ & $-0.036^{+0.035}_{-0.021}$ & - & - &  - &-\\
ESO 264-G036& $\mathbf{0.411^{+0.014}_{-0.017}}$ & $0.428^{+0.017}_{-0.018}$ & $-0.621^{+0.028}_{-0.030}$ & - & - &  - &-\\
ESO 264-G057& $0.675^{+0.022}_{-0.024}$ & $0.423^{+0.022}_{-0.024}$ & $-0.901^{+0.028}_{-0.030}$ & - & - &  - &-\\
ESO 267-G030& $0.535^{+0.017}_{-0.019}$ & $0.353^{+0.013}_{-0.017}$ & $-0.651^{+0.025}_{-0.029}$ & - & - &  - &-\\
ESO 286-G035& $0.669^{+0.031}_{-0.041}$ & $0.344^{+0.009}_{-0.012}$ & $-1.006^{+0.027}_{-0.028}$ & - & - &  - &-\\
ESO 286-IG 019& $\mathbf{0.208^{+0.022}_{-0.022}}$ & $\mathbf{0.175^{+0.03}_{-0.031}}$ & $-3.049^{+0.033}_{-0.033}$ & $\mathbf{3.578^{+0.129}_{-0.099}}$ & $0.936^{+0.009}_{-0.011}$&  -  & II\\
ESO 319-G022& $1.468^{+0.218}_{-0.233}$ & $0.558^{+0.066}_{-0.069}$ & $-0.393^{+0.083}_{-0.078}$ & - & - &  - &-\\
ESO 339-G011& $0.661^{+0.027}_{-0.029}$ & $0.368^{+0.016}_{-0.019}$ & $-0.58^{+0.032}_{-0.032}$ & - & - &  - &-\\
ESO 343-IG013& $0.541^{+0.033}_{-0.033}$ & $0.323^{+0.017}_{-0.019}$ & $-1.25^{+0.040}_{-0.047}$ & $3.330^{+0.679}_{-0.512}$ & $0.537^{+0.043}_{-0.036}$&  - &-\\
ESO 350-IG038& $1.155^{+0.080}_{-0.095}$ & $0.940^{+0.053}_{-0.053}$ & $-0.433^{+0.019}_{-0.021}$ & - & - &  - &-\\
ESO 353-G020& $\mathbf{0.373^{+0.022}_{-0.023}}$ & $0.326^{+0.019}_{-0.02}$ & $-1.561^{+0.061}_{-0.057}$ & $\mathbf{4.589^{+0.467}_{-0.381}}$ & $0.551^{+0.034}_{-0.034}$&  - &-\\
ESO 374-IG032& $\mathbf{0.036^{+0.001}_{-0.001}}$ & $\mathbf{0.102^{+0.007}_{-0.008}}$ & $-4.328^{+0.024}_{-0.023}$ & $\mathbf{6.031^{+0.086}_{-0.081}}$ & $0.959^{+0.001}_{-0.001}$&  -  & I\\
ESO 420- G 013& $\mathbf{0.442^{+0.013}_{-0.015}}$ & $0.353^{+0.009}_{-0.009}$ & $-0.797^{+0.029}_{-0.031}$ & - & - &  - &-\\
ESO 440-IG058& $1.321^{+0.138}_{-0.171}$ & $0.572^{+0.04}_{-0.033}$ & $-0.055^{+0.048}_{-0.03}$ & - & - &  - &-\\
ESO 453-G005& $0.595^{+0.076}_{-0.098}$ & $0.344^{+0.027}_{-0.026}$ & $-0.616^{+0.107}_{-0.116}$ & - & - &  - &-\\
ESO 467-G027& $0.696^{+0.025}_{-0.026}$ & $0.428^{+0.017}_{-0.023}$ & $-0.551^{+0.041}_{-0.045}$ & - & - &  - &-\\
ESO 495- G 021& $2.56^{+0.062}_{-0.072}$ & $0.399^{+0.005}_{-0.005}$ & $-0.151^{+0.010}_{-0.011}$ & - & - &  - &-\\
ESO 507-G070& $0.569^{+0.052}_{-0.060}$ & $\mathbf{0.271^{+0.018}_{-0.021}}$ & $-2.237^{+0.049}_{-0.048}$ & $\mathbf{3.816^{+0.371}_{-0.292}}$ & $0.724^{+0.025}_{-0.025}$&  - &-\\
ESO 557- G 002& $0.698^{+0.048}_{-0.051}$ & $0.296^{+0.02}_{-0.021}$ & $-1.231^{+0.037}_{-0.041}$ & $2.919^{+0.394}_{-0.335}$ & $0.582^{+0.034}_{-0.030}$&  - &-\\
ESO 602-G025& $\mathbf{0.293^{+0.013}_{-0.012}}$ & $\mathbf{0.267^{+0.01}_{-0.011}}$ & $-1.567^{+0.024}_{-0.025}$ & $3.427^{+0.367}_{-0.3}$ & $0.649^{+0.023}_{-0.021}$&  - &-\\
ESO 203-IG001& $\mathbf{0.090^{+0.012}_{-0.014}}$ & $\mathbf{0.173^{+0.027}_{-0.031}}$ & $-4.084^{+0.066}_{-0.061}$ & $\mathbf{5.978^{+0.262}_{-0.235}}$ & $0.943^{+0.005}_{-0.005}$&  - &-\\
ESO 320-G030$^*$& $\mathbf{0.375^{+0.025}_{-0.027}}$ & $\mathbf{0.312^{+0.024}_{-0.026}}$ & $-1.071^{+0.052}_{-0.061}$ & $\mathbf{4.309^{+0.983}_{-0.657}}$ & $0.448^{+0.035}_{-0.033}$&  - &-\\
ESO 557-G002$^*$& $0.858^{+0.075}_{-0.103}$ & $0.451^{+0.036}_{-0.038}$ & $-0.274^{+0.095}_{-0.08}$ & - & - &  - &-\\
ESO 593-IG008& $0.574^{+0.036}_{-0.039}$ & $0.319^{+0.021}_{-0.022}$ & $-1.326^{+0.053}_{-0.057}$ & $2.798^{+0.272}_{-0.237}$ & $0.652^{+0.034}_{-0.036}$&  - &-\\
IC 0214& $0.889^{+0.034}_{-0.039}$ & $0.402^{+0.019}_{-0.023}$ & $-0.726^{+0.034}_{-0.036}$ & - & - &  - &-\\
IC 0563& $0.730^{+0.065}_{-0.072}$ & $0.424^{+0.040}_{-0.042}$ & $-0.809^{+0.084}_{-0.096}$ & - & - &  - &-\\
IC 0860& $\mathbf{0.332^{+0.03}_{-0.032}}$ & $\mathbf{0.187^{+0.016}_{-0.016}}$ & $-1.733^{+0.049}_{-0.054}$ & $\mathbf{4.016^{+0.334}_{-0.281}}$ & $0.758^{+0.016}_{-0.018}$& \checkmark & I\\
IC 2810& $0.722^{+0.062}_{-0.071}$ & $0.380^{+0.031}_{-0.034}$ & $-0.706^{+0.073}_{-0.083}$ & - & - & - &-\\
IC 4280& $0.607^{+0.017}_{-0.019}$ & $0.459^{+0.020}_{-0.019}$ & $-0.592^{+0.037}_{-0.036}$ & - & - & - &-\\
IC 4734& $0.555^{+0.022}_{-0.025}$ & $0.304^{+0.009}_{-0.011}$ & $-1.399^{+0.027}_{-0.029}$ & $2.734^{+0.192}_{-0.172}$ & $0.629^{+0.024}_{-0.027}$& - &-\\
IC 5298& $0.703^{+0.036}_{-0.04}$ & $1.075^{+0.067}_{-0.065}$ & $-0.259^{+0.037}_{-0.036}$ & - & - & - &-\\
IC 4518E$^*$& $1.315^{+0.215}_{-0.278}$ & $0.598^{+0.065}_{-0.061}$ & $-0.536^{+0.137}_{-0.095}$ & - & - & - &-\\
IC 4687$^*$& $1.007^{+0.081}_{-0.086}$ & $0.413^{+0.029}_{-0.032}$ & $-0.648^{+0.067}_{-0.07}$ & - & - & - &-\\
IC 5179$^*$& $0.802^{+0.067}_{-0.083}$ & $0.390^{+0.03}_{-0.034}$ & $-0.839^{+0.069}_{-0.075}$ & - & - & - &-\\
IRAS 00188-0856& $\mathbf{0.143^{+0.016}_{-0.018}}$ & $0.333^{+0.038}_{-0.039}$ & $-2.341^{+0.065}_{-0.06}$ & $3.235^{+0.221}_{-0.19}$ & $0.850^{+0.018}_{-0.019}$& - &-\\
IRAS 00397-1312& $\mathbf{0.179^{+0.025}_{-0.031}}$ & $\mathbf{0.179^{+0.045}_{-0.054}}$ & $-2.940^{+0.043}_{-0.039}$ & $\mathbf{4.059^{+0.222}_{-0.178}}$ & $0.883^{+0.013}_{-0.014}$& - &-\\
IRAS 03521+0028& $0.648^{+0.139}_{-0.215}$ & $0.344^{+0.07}_{-0.085}$ & $-1.207^{+0.158}_{-0.188}$ & $\mathbf{5.090^{+2.356}_{-1.600}}$ & $0.485^{+0.066}_{-0.056}$& -  & II\\
IRAS 04271+3849& $0.577^{+0.015}_{-0.016}$ & $0.357^{+0.014}_{-0.015}$ & $-1.181^{+0.029}_{-0.031}$ & - & - & - &-\\
     \hline
  
   \end{tabular}
   \end{threeparttable}
  \end{table*}

 \begin{table*}
 \ContinuedFloat  
 \centering
   \caption{Spectral Properties Cont.}
     \def\arraystretch{1.2}
    \setlength{\tabcolsep}{1pt}
    \begin{threeparttable}
   \begin{tabular}{cccccccccc}
  
     \hline

    Name & EW(6.2)/EW(11.3) & EW(12.7)/EW(11.3) & $S_{\rm Sil}$ & $\tau_N$ & $\beta$ & HCN (14 $\mu$m) & Crystallines?\\
    (1) & (2) & (3) & (4) & (5) & (6) & (7) & (8)\\
     \hline
IRAS 05083+2441& $0.901^{+0.035}_{-0.036}$ & $0.356^{+0.01}_{-0.011}$ & $-0.958^{+0.033}_{-0.032}$ & - & - & - &-\\
IRAS 05129+5128& $0.856^{+0.045}_{-0.050}$ & $0.342^{+0.022}_{-0.024}$ & $-1.075^{+0.036}_{-0.037}$ & - & - & - &-\\
IRAS 05189-2524& $0.621^{+0.086}_{-0.139}$ & $0.635^{+0.124}_{-0.256}$ & $-0.349^{+0.045}_{-0.063}$ & - & - & - &-\\
IRAS 06035-7102& $\mathbf{0.455^{+0.045}_{-0.045}}$ & $0.348^{+0.055}_{-0.057}$ & $-1.242^{+0.072}_{-0.085}$ & - & - & - &-\\
IRAS 06206-6315& $\mathbf{0.458^{+0.039}_{-0.044}}$ & $0.385^{+0.046}_{-0.045}$ & $-1.562^{+0.065}_{-0.069}$ & - & - & - &-\\
IRAS 07027-6011& $0.736^{+0.043}_{-0.060}$ & $0.475^{+0.022}_{-0.024}$ & $-0.596^{+0.035}_{-0.033}$ & - & - & - &-\\
IRAS 07251-0248& $\mathbf{0.104^{+0.012}_{-0.012}}$ & $\mathbf{0.110^{+0.024}_{-0.025}}$ & $-3.107^{+0.092}_{-0.085}$ & $\mathbf{5.428^{+0.397}_{-0.325}}$ & $0.831^{+0.025}_{-0.020}$& - &-\\
IRAS 07598+6508& $1.128^{+0.293}_{-0.426}$ & $2.608^{+0.554}_{-0.758}$ & $0.147^{+0.016}_{-0.018}$ & - & - & - &-\\
IRAS 08355-4944& $1.019^{+0.029}_{-0.03}$ & $0.774^{+0.033}_{-0.025}$ & $-0.485^{+0.020}_{-0.021}$ & - & - & - &-\\
IRAS 09022-3615& $0.534^{+0.035}_{-0.039}$ & $0.480^{+0.053}_{-0.054}$ & $-1.136^{+0.040}_{-0.043}$ & - & - & - &-\\
IRAS 10378+1109& $\mathbf{0.200^{+0.038}_{-0.049}}$ & $\mathbf{0.225^{+0.072}_{-0.083}}$ & $-2.289^{+0.105}_{-0.113}$ & $\mathbf{4.916^{+0.944}_{-0.679}}$ & $0.721^{+0.039}_{-0.041}$& - &-\\
IRAS 10565+2448& $1.138^{+0.129}_{-0.164}$ & $0.436^{+0.038}_{-0.04}$ & $-0.952^{+0.109}_{-0.116}$ & - & - & - &-\\
IRAS 11095-0238& $\mathbf{0.073^{+0.010}_{-0.011}}$ & $\mathbf{0.011^{+0.006}_{-0.009}}$ & $-3.764^{+0.072}_{-0.066}$ & $\mathbf{5.382^{+0.278}_{-0.242}}$ & $0.925^{+0.009}_{-0.010}$& -  & II\\
IRAS 12116-5615& $\mathbf{0.301^{+0.008}_{-0.009}}$ & $0.276^{+0.006}_{-0.006}$ & $-1.550^{+0.031}_{-0.034}$ & $\mathbf{3.742^{+0.168}_{-0.15}}$ & $0.562^{+0.024}_{-0.025}$& - &-\\
IRAS 13052-5711& $\mathbf{0.478^{+0.026}_{-0.031}}$ & $\mathbf{0.287^{+0.012}_{-0.014}}$ & $-1.438^{+0.032}_{-0.035}$ & $\mathbf{4.121^{+0.501}_{-0.445}}$ & $0.570^{+0.019}_{-0.017}$& - &-\\
IRAS 13120-5453& $0.725^{+0.042}_{-0.047}$ & $0.342^{+0.013}_{-0.015}$ & $-1.554^{+0.045}_{-0.045}$ & - & - & \checkmark&-\\
IRAS 13451+1232& $\mathbf{0.303^{+0.106}_{-0.112}}$ & $0.651^{+0.127}_{-0.131}$ & $-0.254^{+0.018}_{-0.020}$ & - & - & - &-\\
IRAS 14348-1447& $\mathbf{0.279^{+0.028}_{-0.030}}$ & $\mathbf{0.167^{+0.018}_{-0.021}}$ & $-2.066^{+0.06}_{-0.059}$ & $\mathbf{3.909^{+0.452}_{-0.378}}$ & $0.732^{+0.028}_{-0.027}$& - &-\\
IRAS 14378-3651& $0.684^{+0.095}_{-0.122}$ & $0.321^{+0.043}_{-0.052}$ & $-1.750^{+0.094}_{-0.110}$ & $2.642^{+0.381}_{-0.265}$ & $0.709^{+0.052}_{-0.057}$& - &-\\
IRAS 16090-0139& $\mathbf{0.132^{+0.018}_{-0.020}}$ & $\mathbf{0.152^{+0.029}_{-0.033}}$ & $-2.797^{+0.057}_{-0.055}$ & $\mathbf{4.166^{+0.339}_{-0.26}}$ & $0.834^{+0.022}_{-0.025}$& - &-\\
IRAS 17208-0014& $\mathbf{0.331^{+0.028}_{-0.032}}$ & $\mathbf{0.230^{+0.017}_{-0.020}}$ & $-1.97^{+0.086}_{-0.081}$ & $\mathbf{4.805^{+0.429}_{-0.379}}$ & $0.712^{+0.028}_{-0.028}$& -  & I\\
IRAS 17578-0400& $0.643^{+0.054}_{-0.061}$ & $\mathbf{0.270^{+0.014}_{-0.016}}$ & $-1.455^{+0.075}_{-0.073}$ & $\mathbf{4.444^{+0.522}_{-0.443}}$ & $0.513^{+0.040}_{-0.039}$& -  & I\\
IRAS 18090+0130& $0.764^{+0.055}_{-0.066}$ & $0.311^{+0.010}_{-0.012}$ & $-0.76^{+0.041}_{-0.046}$ & $\mathbf{3.603^{+0.595}_{-0.515}}$ & $0.43^{+0.020}_{-0.020}$& - &-\\
IRAS 19254-7245& $\mathbf{0.149^{+0.009}_{-0.01}}$ & $\mathbf{0.203^{+0.024}_{-0.028}}$ & $-1.353^{+0.026}_{-0.026}$ & $\mathbf{4.267^{+0.304}_{-0.273}}$ & $0.432^{+0.013}_{-0.011}$& - &-\\
IRAS 19297-0406& $0.611^{+0.111}_{-0.160}$ & $\mathbf{0.201^{+0.036}_{-0.048}}$ & $-1.680^{+0.133}_{-0.154}$ & $\mathbf{4.746^{+1.369}_{-0.918}}$ & $0.583^{+0.049}_{-0.051}$& - &-\\
IRAS 19542+1110& $0.581^{+0.063}_{-0.072}$ & $0.362^{+0.038}_{-0.039}$ & $-0.899^{+0.097}_{-0.106}$ & - & - & - &-\\
IRAS 20087-0308& $\mathbf{0.267^{+0.032}_{-0.036}}$ & $0.295^{+0.032}_{-0.034}$ & $-2.077^{+0.105}_{-0.109}$ & $\mathbf{4.570^{+0.929}_{-0.659}}$ & $0.668^{+0.034}_{-0.033}$& - &-\\
IRAS 20100-4156& $\mathbf{0.204^{+0.018}_{-0.019}}$ & $\mathbf{0.212^{+0.027}_{-0.030}}$ & $-2.654^{+0.059}_{-0.057}$ & $\mathbf{4.867^{+0.37}_{-0.317}}$ & $0.794^{+0.019}_{-0.019}$& - &-\\
IRAS 20351+2521& $1.027^{+0.069}_{-0.088}$ & $0.491^{+0.022}_{-0.021}$ & $-0.587^{+0.050}_{-0.062}$ & - & - & - &-\\
IRAS 21101+5810& $1.391^{+0.192}_{-0.571}$ & $0.320^{+0.027}_{-0.031}$ & $-0.926^{+0.100}_{-0.157}$ & $2.630^{+0.337}_{-0.281}$ & $0.553^{+0.061}_{-0.073}$& - &-\\
IRAS 22491-1808& $0.647^{+0.09}_{-0.108}$ & $\mathbf{0.179^{+0.024}_{-0.029}}$ & $-1.465^{+0.088}_{-0.088}$ & $\mathbf{4.799^{+0.823}_{-0.626}}$ & $0.652^{+0.036}_{-0.045}$& - &-\\
IRAS 23230-6926& $\mathbf{0.390^{+0.048}_{-0.054}}$ & $\mathbf{0.219^{+0.030}_{-0.031}}$ & $-2.348^{+0.079}_{-0.075}$ & $\mathbf{3.838^{+0.400}_{-0.302}}$ & $0.802^{+0.026}_{-0.027}$& - &-\\
IRAS 23253-5415& $\mathbf{0.432^{+0.047}_{-0.049}}$ & $0.303^{+0.028}_{-0.028}$ & $-1.584^{+0.050}_{-0.056}$ & - & - & - &-\\
IRAS 23365+3604& $0.643^{+0.068}_{-0.082}$ & $\mathbf{0.198^{+0.027}_{-0.030}}$ & $-1.968^{+0.063}_{-0.060}$ & $2.636^{+0.21}_{-0.163}$ & $0.834^{+0.030}_{-0.032}$& - &-\\
IRAS 23436+5257& $0.779^{+0.040}_{-0.044}$ & $0.487^{+0.025}_{-0.026}$ & $-0.506^{+0.030}_{-0.031}$ & - & - & - &-\\
IRAS F01364-1042& $0.605^{+0.108}_{-0.148}$ & $0.301^{+0.035}_{-0.035}$ & $-1.824^{+0.132}_{-0.153}$ & $\mathbf{4.226^{+0.74}_{-0.549}}$ & $0.668^{+0.033}_{-0.033}$& - &-\\
IRAS F01417+1651& $1.387^{+0.449}_{-0.815}$ & $0.302^{+0.036}_{-0.042}$ & $-0.907^{+0.137}_{-0.115}$ & $3.159^{+0.957}_{-0.644}$ & $0.527^{+0.068}_{-0.07}$& - &-\\
IRAS F02437+2122& $\mathbf{0.285^{+0.033}_{-0.036}}$ & $0.286^{+0.037}_{-0.040}$ & $-1.820^{+0.064}_{-0.065}$ & $2.777^{+0.411}_{-0.297}$ & $0.741^{+0.049}_{-0.05}$& - &-\\
IRAS F03217+4022& $0.527^{+0.019}_{-0.023}$ & $0.354^{+0.013}_{-0.014}$ & $-1.238^{+0.030}_{-0.034}$ & - & - & - &-\\
IRAS F05081+7936& $0.828^{+0.048}_{-0.722}$ & $0.462^{+0.026}_{-0.038}$ & $-0.901^{+0.047}_{-0.62}$ & - & - & - &-\\
IRAS F05187-1017& $\mathbf{0.329^{+0.040}_{-0.043}}$ & $0.313^{+0.032}_{-0.035}$ & $-1.741^{+0.116}_{-0.098}$ & $\mathbf{4.292^{+0.995}_{-0.751}}$ & $0.575^{+0.056}_{-0.047}$& - &-\\
IRAS F05189+2524& $0.685^{+0.133}_{-0.128}$ & $0.741^{+0.204}_{-0.259}$ & $-0.320^{+0.065}_{-0.052}$ & - & - & - &-\\
IRAS F06076-2139& $0.596^{+0.054}_{-0.063}$ & $0.376^{+0.034}_{-0.038}$ & $-1.679^{+0.057}_{-0.059}$ & - & - & - &-\\
IRAS F06592-6313& $1.080^{+0.071}_{-0.086}$ & $0.445^{+0.027}_{-0.035}$ & $-0.665^{+0.052}_{-0.065}$ & - & - & - &-\\
IRAS F10173+0828& $\mathbf{0.451^{+0.124}_{-0.215}}$ & $\mathbf{0.199^{+0.049}_{-0.059}}$ & $-1.893^{+0.164}_{-0.194}$ & $\mathbf{5.341^{+1.862}_{-1.254}}$ & $0.669^{+0.050}_{-0.049}$& - &-\\
IRAS F10565+2448& $1.134^{+0.136}_{-0.156}$ & $0.430^{+0.035}_{-0.038}$ & $-0.961^{+0.094}_{-0.099}$ & - & - & - &-\\
IRAS F12112+0305& $\mathbf{0.492^{+0.059}_{-0.060}}$ & $\mathbf{0.232^{+0.028}_{-0.032}}$ & $-1.750^{+0.068}_{-0.078}$ & $2.931^{+0.451}_{-0.294}$ & $0.683^{+0.045}_{-0.048}$& - &-\\
IRAS F12224-0624& $\mathbf{0.063^{+0.010}_{-0.011}}$ & $\mathbf{0.104^{+0.031}_{-0.033}}$ & $-3.500^{+0.089}_{-0.090}$ & $\mathbf{7.129^{+0.609}_{-0.529}}$ & $0.842^{+0.013}_{-0.014}$& - &-\\
IRAS F14348-1447& $\mathbf{0.288^{+0.030}_{-0.033}}$ & $\mathbf{0.172^{+0.020}_{-0.022}}$ & $-2.064^{+0.063}_{-0.070}$ & $\mathbf{3.863^{+0.450}_{-0.348}}$ & $0.729^{+0.027}_{-0.027}$& - &-\\
     \hline
  
   \end{tabular}
   \end{threeparttable}
  \end{table*}

 \begin{table*}
 \ContinuedFloat  
 \centering
   \caption{Spectral Properties Cont.}
     \def\arraystretch{1.2}
    \setlength{\tabcolsep}{1pt}
    \begin{threeparttable}
   \begin{tabular}{cccccccccc}
  
     \hline

    Name & EW(6.2)/EW(11.3) & EW(12.7)/EW(11.3) & $S_{\rm Sil}$ & $\tau_N$ & $\beta$ & HCN (14 $\mu$m) & Crystallines?\\
    (1) & (2) & (3) & (4) & (5) & (6) & (7) & (8)\\
     \hline
IRAS F14378-3651& $0.688^{+0.095}_{-0.129}$ & $0.324^{+0.045}_{-0.050}$ & $-1.752^{+0.097}_{-0.107}$ & $2.625^{+0.374}_{-0.262}$ & $0.712^{+0.053}_{-0.062}$& - &-\\
IRAS F16164-0746& $0.556^{+0.066}_{-0.072}$ & $0.335^{+0.025}_{-0.023}$ & $-2.097^{+0.07}_{-0.069}$ & $\mathbf{4.836^{+0.657}_{-0.533}}$ & $0.652^{+0.024}_{-0.023}$& - &-\\
IRAS F16516-0948& $0.760^{+0.041}_{-0.049}$ & $0.410^{+0.027}_{-0.028}$ & $-0.628^{+0.058}_{-0.066}$ & - & - & - &-\\
IRAS F17138-1017& $0.613^{+0.035}_{-0.044}$ & $0.352^{+0.01}_{-0.012}$ & $-1.106^{+0.037}_{-0.041}$ & - & - & - &-\\
IRAS F17207-0014& $\mathbf{0.331^{+0.027}_{-0.03}}$ & $\mathbf{0.231^{+0.017}_{-0.019}}$ & $-1.968^{+0.074}_{-0.069}$ & $\mathbf{4.804^{+0.429}_{-0.382}}$ & $0.711^{+0.027}_{-0.029}$& - &-\\
IRAS F18293-3413& $0.600^{+0.014}_{-0.014}$ & $0.318^{+0.007}_{-0.008}$ & $-1.533^{+0.014}_{-0.016}$ & - & - & - &-\\
IRAS F19297-0406& $0.567^{+0.107}_{-0.148}$ & $\mathbf{0.191^{+0.035}_{-0.045}}$ & $-1.697^{+0.119}_{-0.143}$ & $\mathbf{4.690^{+1.381}_{-0.848}}$ & $0.607^{+0.048}_{-0.055}$& - &-\\
IRAS F22491-1808& $0.652^{+0.084}_{-0.101}$ & $\mathbf{0.178^{+0.024}_{-0.027}}$ & $-1.470^{+0.071}_{-0.078}$ & $\mathbf{4.767^{+0.853}_{-0.647}}$ & $0.656^{+0.035}_{-0.051}$& - &-\\
IRAS F23365+3604& $0.667^{+0.07}_{-0.082}$ & $\mathbf{0.207^{+0.029}_{-0.032}}$ & $-1.964^{+0.049}_{-0.057}$ & $2.605^{+0.201}_{-0.163}$ & $0.834^{+0.033}_{-0.035}$& - &-\\
M-3-34-63& $0.588^{+0.209}_{-0.299}$ & $0.419^{+0.127}_{-0.16}$ & $-0.477^{+0.128}_{-0.149}$ & - & - & - &-\\
MCG +08-11-002& $\mathbf{0.336^{+0.017}_{-0.017}}$ & $\mathbf{0.232^{+0.008}_{-0.009}}$ & $-2.380^{+0.046}_{-0.039}$ & $\mathbf{4.705^{+0.219}_{-0.195}}$ & $0.716^{+0.016}_{-0.013}$& - &-\\
MCG +09-27-025& $0.749^{+0.191}_{-0.261}$ & $0.499^{+0.071}_{-0.084}$ & $-0.831^{+0.245}_{-0.126}$ & - & - & - &-\\
MCG +10-25-031& $0.581^{+0.077}_{-0.098}$ & $0.446^{+0.042}_{-0.049}$ & $-0.247^{+0.093}_{-0.077}$ & - & - & - &-\\
MCG -02-33-098& $1.108^{+0.065}_{-0.076}$ & $0.435^{+0.019}_{-0.025}$ & $-0.507^{+0.047}_{-0.047}$ & - & - & - &-\\
MCG -05-12-006& $1.592^{+0.104}_{-0.12}$ & $0.452^{+0.021}_{-0.031}$ & $-0.246^{+0.033}_{-0.031}$ & - & - & - &-\\
MCG+04-48-002& $\mathbf{0.421^{+0.018}_{-0.018}}$ & $0.334^{+0.015}_{-0.018}$ & $-1.228^{+0.048}_{-0.051}$ & $3.403^{+0.506}_{-0.389}$ & $0.403^{+0.053}_{-0.052}$& - &-\\
MCG+12-02-001& $1.293^{+0.032}_{-0.032}$ & $0.527^{+0.01}_{-0.015}$ & $-0.579^{+0.015}_{-0.015}$ & - & - & - &-\\
MCG-02-01-051A& $0.774^{+0.121}_{-0.136}$ & $0.403^{+0.063}_{-0.068}$ & $-0.419^{+0.084}_{-0.117}$ & - & - & - &-\\
MCG-02-33-098& $1.691^{+0.05}_{-0.052}$ & $0.694^{+0.039}_{-0.038}$ & $-0.229^{+0.018}_{-0.018}$ & - & - & - &-\\
MCG-03-04-014& $1.018^{+0.054}_{-0.063}$ & $0.444^{+0.026}_{-0.031}$ & $-0.684^{+0.049}_{-0.055}$ & - & - & - &-\\
MCG-03-34-064& $0.191^{+0.034}_{-0.036}$ & $0.526^{+0.04}_{-0.041}$ & $-0.091^{+0.004}_{-0.004}$ & - & - & - &-\\
Mrk 0231& $0.770^{+0.088}_{-0.097}$ & $0.547^{+0.122}_{-0.141}$ & $-0.689^{+0.02}_{-0.022}$ & - & - & - &-\\
Mrk 0273& $\mathbf{0.409^{+0.04}_{-0.048}}$ & $\mathbf{0.254^{+0.035}_{-0.039}}$ & $-2.061^{+0.039}_{-0.038}$ & $2.883^{+0.2}_{-0.175}$ & $0.796^{+0.03}_{-0.031}$& - &-\\
Mrk 0331& $0.592^{+0.025}_{-0.026}$ & $0.297^{+0.011}_{-0.013}$ & $-1.030^{+0.022}_{-0.023}$ & $2.725^{+0.324}_{-0.285}$ & $0.55^{+0.029}_{-0.024}$& - &-\\
Mrk 1014& $0.841^{+0.127}_{-0.153}$ & $0.644^{+0.122}_{-0.144}$ & $0.205^{+0.03}_{-0.033}$ & - & - & - &-\\
Mrk 1490& $1.116^{+0.077}_{-0.084}$ & $0.414^{+0.032}_{-0.034}$ & $-0.682^{+0.043}_{-0.047}$ & - & - & - &-\\
Mrk 938$^*$& $\mathbf{0.499^{+0.043}_{-0.048}}$ & $0.337^{+0.035}_{-0.037}$ & $-1.569^{+0.06}_{-0.062}$ & $2.623^{+0.285}_{-0.207}$ & $0.639^{+0.041}_{-0.042}$& - &-\\
NGC 0023& $0.665^{+0.019}_{-0.021}$ & $0.418^{+0.014}_{-0.017}$ & $-0.448^{+0.028}_{-0.031}$ & - & - & - &-\\
NGC 0232& $0.545^{+0.015}_{-0.016}$ & $0.335^{+0.006}_{-0.007}$ & $-1.021^{+0.021}_{-0.022}$ & $2.606^{+0.238}_{-0.211}$ & $0.476^{+0.027}_{-0.024}$& - &-\\
NGC 0354& $0.683^{+0.025}_{-0.03}$ & $0.420^{+0.021}_{-0.023}$ & $-0.709^{+0.048}_{-0.049}$ & - & - & - &-\\
NGC 0633& $0.734^{+0.015}_{-0.017}$ & $0.487^{+0.023}_{-0.033}$ & $-0.505^{+0.033}_{-0.036}$ & - & - & - &-\\
NGC 0660& $0.582^{+0.025}_{-0.028}$ & $0.289^{+0.007}_{-0.008}$ & $-1.659^{+0.026}_{-0.027}$ & $2.798^{+0.16}_{-0.14}$ & $0.665^{+0.024}_{-0.027}$& - &-\\
NGC 0695& $0.842^{+0.028}_{-0.032}$ & $0.444^{+0.023}_{-0.024}$ & $-0.613^{+0.033}_{-0.035}$ & - & - & - &-\\
NGC 0828& $\mathbf{0.459^{+0.01}_{-0.01}}$ & $0.350^{+0.009}_{-0.009}$ & $-1.144^{+0.014}_{-0.014}$ & - & - & - &-\\
NGC 0838& $1.292^{+0.196}_{-0.207}$ & $0.428^{+0.03}_{-0.028}$ & $-0.266^{+0.098}_{-0.053}$ & - & - & - &-\\
NGC 0877& $\mathbf{0.275^{+0.017}_{-0.021}}$ & $0.464^{+0.024}_{-0.024}$ & $-0.463^{+0.044}_{-0.051}$ & - & - & - &-\\
NGC 0958& $\mathbf{0.380^{+0.017}_{-0.02}}$ & $0.497^{+0.033}_{-0.032}$ & $-0.526^{+0.055}_{-0.058}$ & - & - & - &-\\
NGC 0992& $0.844^{+0.018}_{-0.019}$ & $0.378^{+0.009}_{-0.011}$ & $-0.769^{+0.022}_{-0.023}$ & - & - & - &-\\
NGC 1067& $0.949^{+0.089}_{-0.103}$ & $0.444^{+0.041}_{-0.043}$ & $-0.192^{+0.076}_{-0.053}$ & - & - & - &-\\
NGC 1365& $\mathbf{0.419^{+0.01}_{-0.011}}$ & $0.520^{+0.013}_{-0.016}$ & $-0.183^{+0.021}_{-0.023}$ & - & - & - &-\\
NGC 1572& $1.202^{+0.049}_{-0.05}$ & $0.621^{+0.034}_{-0.033}$ & $-0.327^{+0.044}_{-0.04}$ & - & - & - &-\\
NGC 1614& $1.471^{+0.038}_{-0.043}$ & $0.472^{+0.02}_{-0.024}$ & $-0.698^{+0.015}_{-0.016}$ & - & - & - &-\\
NGC 1797& $0.942^{+0.043}_{-0.049}$ & $0.465^{+0.027}_{-0.033}$ & $-0.712^{+0.03}_{-0.033}$ & - & - & - &-\\
NGC 1808& $0.696^{+0.013}_{-0.013}$ & $0.393^{+0.009}_{-0.01}$ & $-0.796^{+0.021}_{-0.027}$ & - & - & - &-\\
NGC 1961& $\mathbf{0.147^{+0.01}_{-0.011}}$ & $0.489^{+0.028}_{-0.026}$ & $-0.518^{+0.051}_{-0.046}$ & - & - & - &-\\
NGC 2146& $1.417^{+0.078}_{-0.089}$ & $0.392^{+0.011}_{-0.01}$ & $-0.707^{+0.02}_{-0.023}$ & - & - & - &-\\
NGC 2342 A$^*$& $1.077^{+0.063}_{-0.072}$ & $0.510^{+0.036}_{-0.032}$ & $-0.321^{+0.079}_{-0.074}$ & - & - & - &-\\
NGC 2342& $0.769^{+0.067}_{-0.078}$ & $0.377^{+0.02}_{-0.025}$ & $-0.660^{+0.053}_{-0.059}$ & - & - & - &-\\
NGC 2369$^*$& $0.654^{+0.031}_{-0.03}$ & $0.367^{+0.022}_{-0.028}$ & $-1.329^{+0.041}_{-0.04}$ & - & - & - &-\\
     \hline
  
   \end{tabular}
   \end{threeparttable}
  \end{table*}

 \begin{table*}
 \ContinuedFloat  
 \centering
   \caption{Spectral Properties Cont.}
     \def\arraystretch{1.2}
    \setlength{\tabcolsep}{1pt}
    \begin{threeparttable}
   \begin{tabular}{cccccccccc}
  
     \hline

    Name & EW(6.2)/EW(11.3) & EW(12.7)/EW(11.3) & $S_{\rm Sil}$ & $\tau_N$ & $\beta$ & HCN (14 $\mu$m) & Crystallines?\\
    (1) & (2) & (3) & (4) & (5) & (6) & (7) & (8)\\
     \hline
NGC 2388 A$^*$& $0.661^{+0.055}_{-0.063}$ & $0.474^{+0.041}_{-0.048}$ & $-0.346^{+0.095}_{-0.098}$ & - & - & - &-\\
NGC 2388& $0.750^{+0.017}_{-0.019}$ & $0.406^{+0.009}_{-0.011}$ & $-0.778^{+0.021}_{-0.023}$ & - & - & - &-\\
NGC 2544& $0.503^{+0.030}_{-0.036}$ & $0.387^{+0.031}_{-0.036}$ & $-0.479^{+0.055}_{-0.065}$ & - & - & - &-\\
NGC 2623& $0.690^{+0.079}_{-0.105}$ & $0.329^{+0.028}_{-0.028}$ & $-1.860^{+0.073}_{-0.079}$ & $3.065^{+0.302}_{-0.254}$ & $0.731^{+0.032}_{-0.032}$& - &-\\
NGC 2903& $0.877^{+0.024}_{-0.026}$ & $0.530^{+0.013}_{-0.011}$ & $-0.310^{+0.023}_{-0.02}$ & - & - & - &-\\
NGC 2993& $1.529^{+0.093}_{-0.138}$ & $0.514^{+0.025}_{-0.025}$ & $-0.021^{+0.022}_{-0.012}$ & - & - & - &-\\
NGC 3110& $0.833^{+0.03}_{-0.034}$ & $0.401^{+0.013}_{-0.015}$ & $-0.834^{+0.036}_{-0.043}$ & - & - & - &-\\
NGC 3188& $1.099^{+0.123}_{-0.141}$ & $0.464^{+0.05}_{-0.053}$ & $-0.147^{+0.087}_{-0.071}$ & - & - & - &-\\
NGC 3256& $1.037^{+0.031}_{-0.037}$ & $0.396^{+0.009}_{-0.009}$ & $-0.647^{+0.022}_{-0.022}$ & - & - & - &-\\
NGC 3628& $\mathbf{0.378^{+0.023}_{-0.024}}$ & $0.283^{+0.015}_{-0.016}$ & $-2.360^{+0.04}_{-0.041}$ & $\mathbf{5.098^{+0.495}_{-0.401}}$ & $0.669^{+0.014}_{-0.014}$& - &-\\
NGC 4194& $0.953^{+0.027}_{-0.03}$ & $0.433^{+0.011}_{-0.014}$ & $-0.723^{+0.02}_{-0.021}$ & - & - & - &-\\
NGC 4369& $0.643^{+0.017}_{-0.019}$ & $0.392^{+0.013}_{-0.015}$ & $-0.362^{+0.026}_{-0.026}$ & - & - & - &-\\
NGC 4385& $1.921^{+0.12}_{-0.196}$ & $0.546^{+0.043}_{-0.034}$ & $-0.079^{+0.047}_{-0.035}$ & - & - & - &-\\
NGC 4418& $\mathbf{0.038^{+0.005}_{-0.005}}$ & $\mathbf{0.096^{+0.023}_{-0.024}}$ & $-4.317^{+0.019}_{-0.018}$ & $\mathbf{4.546^{+0.037}_{-0.035}}$ & $0.988^{+0.001}_{-0.002}$ & \checkmark & I \\
NGC 4666& $\mathbf{0.458^{+0.034}_{-0.032}}$ & $0.527^{+0.025}_{-0.022}$ & $-0.333^{+0.084}_{-0.082}$ & - & - & - &-\\
NGC 4922$^*$& $0.617^{+0.07}_{-0.078}$ & $0.530^{+0.057}_{-0.061}$ & $-0.519^{+0.034}_{-0.02}$ & - & - & - &-\\
NGC 4922& $1.021^{+0.096}_{-0.114}$ & $0.480^{+0.057}_{-0.059}$ & $-0.515^{+0.064}_{-0.045}$ & - & - & - &-\\
NGC 5010& $\mathbf{0.326^{+0.013}_{-0.014}}$ & $0.291^{+0.012}_{-0.015}$ & $-1.494^{+0.03}_{-0.036}$ & $\mathbf{3.651^{+0.318}_{-0.292}}$ & $0.616^{+0.019}_{-0.019}$& - &-\\
NGC 5104& $\mathbf{0.388^{+0.027}_{-0.027}}$ & $0.302^{+0.022}_{-0.025}$ & $-1.298^{+0.049}_{-0.081}$ & $3.186^{+0.404}_{-0.311}$ & $0.590^{+0.038}_{-0.04}$& - &-\\
NGC 5135& $0.523^{+0.024}_{-0.019}$ & $0.456^{+0.023}_{-0.022}$ & $-0.717^{+0.029}_{-0.031}$ & - & - & - &-\\
NGC 5257 A$^*$& $0.683^{+0.053}_{-0.056}$ & $0.504^{+0.046}_{-0.048}$ & $-0.455^{+0.073}_{-0.089}$ & - & - & - &-\\
NGC 5331& $0.614^{+0.025}_{-0.027}$ & $0.396^{+0.026}_{-0.024}$ & $-1.136^{+0.038}_{-0.038}$ & - & - & - &-\\
NGC 5394& $0.687^{+0.027}_{-0.03}$ & $0.358^{+0.014}_{-0.016}$ & $-0.648^{+0.023}_{-0.025}$ & - & - & - &-\\
NGC 5395$^*$& $\mathbf{0.125^{+0.016}_{-0.017}}$ & $0.408^{+0.037}_{-0.04}$ & $-0.355^{+0.026}_{-0.024}$ & - & - & - &-\\
NGC 5430& $1.040^{+0.037}_{-0.041}$ & $0.612^{+0.012}_{-0.01}$ & $-0.263^{+0.018}_{-0.02}$ & - & - & - &-\\
NGC 5607& $1.015^{+0.036}_{-0.039}$ & $0.443^{+0.018}_{-0.02}$ & $-0.522^{+0.026}_{-0.028}$ & - & - & - &-\\
NGC 5643& $0.611^{+0.032}_{-0.034}$ & $0.423^{+0.02}_{-0.019}$ & $-0.309^{+0.032}_{-0.033}$ & - & - & - &-\\
NGC 5734& $\mathbf{0.396^{+0.017}_{-0.019}}$ & $0.391^{+0.017}_{-0.02}$ & $-0.603^{+0.027}_{-0.029}$ & - & - & - &-\\
NGC 5743& $\mathbf{0.397^{+0.023}_{-0.026}}$ & $0.323^{+0.02}_{-0.026}$ & $-0.804^{+0.055}_{-0.059}$ & - & - & - &-\\
NGC 5936& $0.752^{+0.044}_{-0.042}$ & $0.392^{+0.019}_{-0.021}$ & $-0.742^{+0.031}_{-0.033}$ & - & - & - &-\\
NGC 5990& $\mathbf{0.254^{+0.005}_{-0.006}}$ & $0.448^{+0.018}_{-0.019}$ & $-0.639^{+0.024}_{-0.023}$ & - & - & - &-\\
NGC 6090 NED01& $2.174^{+0.205}_{-0.287}$ & $0.397^{+0.02}_{-0.021}$ & $-0.145^{+0.038}_{-0.028}$ & - & - & - &-\\
NGC 6090& $1.943^{+0.222}_{-0.271}$ & $0.361^{+0.018}_{-0.02}$ & $-0.207^{+0.03}_{-0.025}$ & - & - & - &-\\
NGC 6161& $1.255^{+0.376}_{-0.317}$ & $0.456^{+0.06}_{-0.053}$ & $-0.663^{+0.277}_{-0.131}$ & - & - & - &-\\
NGC 6240& $0.820^{+0.042}_{-0.044}$ & $0.339^{+0.016}_{-0.013}$ & $-1.438^{+0.038}_{-0.034}$ & - & - & - &-\\
NGC 6286& $\mathbf{0.389^{+0.018}_{-0.019}}$ & $0.274^{+0.011}_{-0.012}$ & $-1.557^{+0.026}_{-0.03}$ & $\mathbf{3.939^{+0.384}_{-0.349}}$ & $0.628^{+0.018}_{-0.017}$& - &-\\
NGC 6670A& $0.867^{+0.036}_{-0.038}$ & $0.350^{+0.014}_{-0.017}$ & $-0.844^{+0.037}_{-0.039}$ & - & - & - &-\\
NGC 6701& $0.520^{+0.026}_{-0.025}$ & $0.359^{+0.017}_{-0.017}$ & $-0.807^{+0.025}_{-0.026}$ & - & - & - &-\\
NGC 6786 A$^*$& $0.806^{+0.034}_{-0.036}$ & $0.528^{+0.03}_{-0.026}$ & $-0.383^{+0.016}_{-0.016}$ & - & - & - &-\\
NGC 6786& $0.931^{+0.049}_{-0.048}$ & $0.521^{+0.028}_{-0.024}$ & $-0.348^{+0.056}_{-0.054}$ & - & - & - &-\\
NGC 6926$^*$& $\mathbf{0.343^{+0.028}_{-0.031}}$ & $0.365^{+0.028}_{-0.029}$ & $-1.120^{+0.049}_{-0.052}$ & - & - & - &-\\
NGC 6926& $\mathbf{0.280^{+0.062}_{-0.1}}$ & $\mathbf{0.258^{+0.056}_{-0.067}}$ & $-1.352^{+0.195}_{-0.176}$ & $\mathbf{5.633^{+2.133}_{-1.465}}$ & $0.502^{+0.054}_{-0.061}$& - &-\\
NGC 7130& $0.792^{+0.042}_{-0.043}$ & $0.507^{+0.026}_{-0.023}$ & $-0.462^{+0.031}_{-0.037}$ & - & - & - &-\\
NGC 7252& $0.658^{+0.032}_{-0.039}$ & $0.419^{+0.021}_{-0.024}$ & $-0.474^{+0.05}_{-0.056}$ & - & - & - &-\\
NGC 7469& $0.975^{+0.017}_{-0.018}$ & $0.451^{+0.006}_{-0.006}$ & $-0.053^{+0.002}_{-0.002}$ & - & - & - &-\\
NGC 7552$^*$& $0.595^{+0.039}_{-0.041}$ & $0.376^{+0.026}_{-0.027}$ & $-0.735^{+0.068}_{-0.076}$ & - & - & - &-\\
NGC 7591& $0.521^{+0.074}_{-0.081}$ & $0.354^{+0.032}_{-0.037}$ & $-1.038^{+0.096}_{-0.1}$ & $2.818^{+1.159}_{-0.695}$ & $0.409^{+0.104}_{-0.092}$& - &-\\
NGC 7592 A$^*$& $0.956^{+0.06}_{-0.062}$ & $0.445^{+0.025}_{-0.027}$ & $-0.745^{+0.044}_{-0.047}$ & - & - & - &-\\
NGC 7592 B$^*$& $\mathbf{0.468^{+0.017}_{-0.016}}$ & $0.458^{+0.018}_{-0.018}$ & $-1.446^{+0.024}_{-0.024}$ & - & - & - &-\\
     \hline
  
   \end{tabular}
   \end{threeparttable}
  \end{table*}

 \begin{table*}
 \ContinuedFloat  
 \centering
   \caption{Spectral Properties Cont.}
     \def\arraystretch{1.2}
    \setlength{\tabcolsep}{1pt}
    \begin{threeparttable}
   \begin{tabular}{cccccccccc}
  
     \hline

    Name & EW(6.2)/EW(11.3) & EW(12.7)/EW(11.3) & $S_{\rm Sil}$ & $\tau_N$ & $\beta$ & HCN (14 $\mu$m) & Crystallines?\\
    (1) & (2) & (3) & (4) & (5) & (6) & (7) & (8)\\
     \hline
NGC 7674$^*$& $\mathbf{0.243^{+0.057}_{-0.061}}$ & $0.442^{+0.062}_{-0.065}$ & $-0.185^{+0.012}_{-0.012}$ & - & - & \checkmark &-\\
NGC 7674A& $0.920^{+0.078}_{-0.055}$ & $0.485^{+0.036}_{-0.04}$ & $-0.35^{+0.077}_{-0.082}$ & - & - & - &-\\
NGC 7679& $0.962^{+0.017}_{-0.016}$ & $0.462^{+0.008}_{-0.009}$ & $-0.195^{+0.014}_{-0.015}$ & - & - & - &-\\
NGC 7752 A$^*$& $\mathbf{0.326^{+0.023}_{-0.024}}$ & $0.544^{+0.033}_{-0.034}$ & $-0.500^{+0.037}_{-0.037}$ & - & - & - &-\\
NGC 7752& $0.704^{+0.053}_{-0.064}$ & $0.365^{+0.024}_{-0.029}$ & $-0.698^{+0.051}_{-0.056}$ & - & - & - &-\\
NGC 7771& $0.603^{+0.03}_{-0.037}$ & $0.427^{+0.018}_{-0.017}$ & $-0.823^{+0.046}_{-0.049}$ & $2.875^{+0.635}_{-0.505}$ & $0.401^{+0.049}_{-0.039}$& - &-\\
NVSS J211129+582307& $1.388^{+0.191}_{-0.563}$ & $0.319^{+0.027}_{-0.03}$ & $-0.932^{+0.09}_{-0.152}$ & $2.586^{+0.342}_{-0.267}$ & $0.562^{+0.056}_{-0.075}$& - &-\\
SBS 1132+579& $0.817^{+0.078}_{-0.083}$ & $0.559^{+0.046}_{-0.042}$ & $-0.212^{+0.114}_{-0.089}$ & - & - & - &-\\
SBS 1204+579& $0.789^{+0.062}_{-0.062}$ & $0.525^{+0.057}_{-0.06}$ & $-0.228^{+0.104}_{-0.081}$ & - & - & - &-\\
UGC 01385& $2.120^{+0.153}_{-0.169}$ & $0.438^{+0.025}_{-0.025}$ & $-0.185^{+0.028}_{-0.025}$ & - & -   & - &-\\
UGC 01845& $\mathbf{0.390^{+0.016}_{-0.014}}$ & $\mathbf{0.271^{+0.012}_{-0.012}}$ & $-1.630^{+0.017}_{-0.018}$ & $\mathbf{3.501^{+0.341}_{-0.234}}$ & $0.685^{+0.015}_{-0.016}$  & \checkmark &-\\
UGC 02238& $\mathbf{0.419^{+0.027}_{-0.029}}$ & $0.280^{+0.014}_{-0.015}$ & $-1.556^{+0.059}_{-0.062}$ & $3.340^{+0.377}_{-0.311}$ & $0.643^{+0.026}_{-0.031}$& - &-\\
UGC 02608& $0.507^{+0.039}_{-0.041}$ & $0.409^{+0.032}_{-0.036}$ & $-0.767^{+0.055}_{-0.062}$ & - & - & - &-\\
UGC 02894& $0.856^{+0.051}_{-0.056}$ & $0.445^{+0.023}_{-0.021}$ & $-0.422^{+0.067}_{-0.064}$ & - & - & - &-\\
UGC 02982& $0.589^{+0.01}_{-0.01}$ & $0.373^{+0.01}_{-0.011}$ & $-0.947^{+0.022}_{-0.022}$ & - & - & - &-\\
UGC 03094& $0.509^{+0.02}_{-0.021}$ & $0.349^{+0.013}_{-0.016}$ & $-0.820^{+0.037}_{-0.042}$ & $2.617^{+0.527}_{-0.39}$ & $0.406^{+0.046}_{-0.041}$& - &-\\
UGC 03351& $\mathbf{0.376^{+0.016}_{-0.016}}$ & $0.301^{+0.012}_{-0.012}$ & $-1.567^{+0.046}_{-0.045}$ & $\mathbf{3.636^{+0.239}_{-0.205}}$ & $0.621^{+0.025}_{-0.025}$& - &-\\
UGC 03356 NOTES01& $0.847^{+0.032}_{-0.034}$ & $0.379^{+0.014}_{-0.015}$ & $-1.033^{+0.025}_{-0.026}$ & - & - & - &-\\
UGC 03405& $\mathbf{0.473^{+0.033}_{-0.038}}$ & $0.373^{+0.028}_{-0.028}$ & $-0.785^{+0.057}_{-0.062}$ & - & - & - &-\\
UGC 03410$^*$& $0.745^{+0.09}_{-0.097}$ & $0.446^{+0.046}_{-0.047}$ & $-0.654^{+0.113}_{-0.116}$ & - & - & - &-\\
UGC 03410& $0.530^{+0.023}_{-0.023}$ & $0.356^{+0.017}_{-0.015}$ & $-0.946^{+0.033}_{-0.035}$ & $2.933^{+0.55}_{-0.392}$ & $0.472^{+0.036}_{-0.033}$& - &-\\
UGC 03608& $1.293^{+0.069}_{-0.077}$ & $0.384^{+0.02}_{-0.019}$ & $-0.455^{+0.034}_{-0.034}$ & - & - & - &-\\
UGC 04261& $1.902^{+0.173}_{-0.346}$ & $0.413^{+0.022}_{-0.025}$ & $-0.030^{+0.029}_{-0.017}$ & - & - & - &-\\
UGC 04438& $0.890^{+0.079}_{-0.094}$ & $0.479^{+0.037}_{-0.044}$ & $-0.105^{+0.062}_{-0.041}$ & - & - & - &-\\
UGC 04881& $0.549^{+0.034}_{-0.042}$ & $0.389^{+0.021}_{-0.02}$ & $-1.110^{+0.056}_{-0.064}$ & - & - & - &-\\
UGC 05101& $0.269^{+0.023}_{-0.027}$ & $0.480^{+0.041}_{-0.044}$ & $-1.444^{+0.053}_{-0.053}$ & - & - & - &-\\
UGC 05408& $1.572^{+0.157}_{-0.168}$ & $0.390^{+0.026}_{-0.029}$ & $-0.130^{+0.043}_{-0.03}$ & - & - & - &-\\
UGC 06514 NED01& $0.568^{+0.082}_{-0.123}$ & $0.539^{+0.047}_{-0.049}$ & $-0.395^{+0.113}_{-0.101}$ & - & - & - &-\\
UGC 08335 NED02& $1.276^{+0.073}_{-0.085}$ & $0.403^{+0.027}_{-0.031}$ & $-1.067^{+0.043}_{-0.048}$ & - & - & - &-\\
UGC 08335& $1.053^{+0.25}_{-0.257}$ & $0.371^{+0.032}_{-0.038}$ & $-0.533^{+0.175}_{-0.118}$ & - & - & - &-\\
UGC 08739& $\mathbf{0.281^{+0.021}_{-0.021}}$ & $0.296^{+0.02}_{-0.021}$ & $-1.735^{+0.069}_{-0.061}$ & $\mathbf{4.326^{+0.492}_{-0.407}}$ & $0.595^{+0.035}_{-0.036}$& - &-\\
UGC 08850& $\mathbf{0.099^{+0.026}_{-0.027}}$ & $0.375^{+0.051}_{-0.053}$ & $-0.388^{+0.009}_{-0.009}$ & - & - & - &-\\
UGC 11041& $0.510^{+0.019}_{-0.019}$ & $0.403^{+0.017}_{-0.018}$ & $-0.659^{+0.037}_{-0.039}$ & - & - & - &-\\
UGC 12150& $0.562^{+0.026}_{-0.027}$ & $0.318^{+0.01}_{-0.011}$ & $-1.172^{+0.032}_{-0.033}$ & - & - & - &-\\
UGC 2982& $1.280^{+0.048}_{-0.054}$ & $0.445^{+0.013}_{-0.014}$ & $-0.205^{+0.022}_{-0.016}$ & - & - & - &-\\
UGC 8387& $0.572^{+0.048}_{-0.049}$ & $\mathbf{0.259^{+0.013}_{-0.014}}$ & $-1.924^{+0.068}_{-0.063}$ & $\mathbf{3.590^{+0.285}_{-0.237}}$ & $0.705^{+0.027}_{-0.03}$& - &-\\
VII Zw 031& $1.544^{+0.102}_{-0.119}$ & $0.500^{+0.016}_{-0.016}$ & $-0.284^{+0.036}_{-0.031}$ & - & - & - &-\\
VV 059a& $\mathbf{0.486^{+0.035}_{-0.041}}$ & $0.303^{+0.014}_{-0.017}$ & $-1.392^{+0.051}_{-0.057}$ & $\mathbf{5.705^{+1.258}_{-0.954}}$ & $0.533^{+0.025}_{-0.027}$& - &-\\
VV 283a& $\mathbf{0.476^{+0.04}_{-0.043}}$ & $\mathbf{0.267^{+0.014}_{-0.016}}$ & $-1.877^{+0.07}_{-0.068}$ & $\mathbf{3.908^{+0.322}_{-0.287}}$ & $0.700^{+0.024}_{-0.025}$& - &-\\
ZW049.057$^*$& $\mathbf{0.459^{+0.058}_{-0.063}}$ & $\mathbf{0.278^{+0.034}_{-0.036}}$ & $-1.679^{+0.081}_{-0.085}$ & $\mathbf{4.444^{+0.962}_{-0.697}}$ & $0.519^{+0.05}_{-0.044}$& - &-\\

    \hline
  
  \end{tabular}
\begin{tablenotes}
    \item[$\bullet$] Column (1): Object Name. Column (2): Ratio of the EW of the 6.2 $\mu$m PAH to the 11.3 $\mu$m PAH. Column (3): Ratio of the EW of the 12.7 $\mu$m PAH to the 11.3 $\mu$m PAH. Column (4): Silicate strength according to equnation (\ref{eqn:SilStrength}). Column (5): Nuclear optical depth. Column (6): Nuclear fraction. Column (7): Detection of HCN (14 $\mu$m) absorption above 5$\sigma$ in low-res Spitzer spectra. Column (8): Detection of Crystalline silicate absorption from \citet{IDEOS}. Method I selects sources with $z<0.068$, $s_{23}<0$ and $s_{33}<0$. Method II selects sources with $z<0.257$, $s_{23}<-0.09$ and $s_{33}<-0.02$. 
    \item[$\bullet$] Sources marked $^*$ use the innermost aperture from the spectral maps
    \item[$\bullet$] Values for $\tau_N$ and $\beta$ are only shown for $\beta > 0.4$ and $\tau_N>2.5$ where these values are reliable as discussed in Section \ref{sec:Testing}. 
    \item[$\bullet$] PAH EW ratios with values lower than the threshold for CON classification are shown in bold. For those from spectral maps marked as $^*$, the threshold used is the one adjusted for the aperture correction. Optical depths of $\tau_N>3.5$ are also shown in bold.
  \end{tablenotes}
  \end{threeparttable}
 \end{table*}

\end{document}